%% file: GB_CausalityD_30.tex
\documentclass[11pt,a4paper]{article}
\usepackage{jheppub}
\usepackage[utf8]{inputenc}

\usepackage{mathtools, tensor}
\usepackage[dvipsnames]{xcolor}
\usepackage{enumitem}
\usepackage{microtype}

\newcommand{\RGB}{R_{\text{GB}}}                
\newcommand{\cGB}{c_{\text{GB}}}                
\newcommand{\rgr}{\left(\frac{r_g}{r}\right)}   
\newcommand{\Mpl}{M_{\rm Pl}}                
\newcommand{\Scalar}{\mathbb{S}}                
\newcommand{\VV}{\mathbb{V}}                    
\newcommand{\TT}{\mathbb{T}}                    
\newcommand{\LB}{\hat{\Delta}_{D-2}}            

\DeclarePairedDelimiter\abs{\lvert}{\rvert}     

\renewcommand{\(}{\left(}
\renewcommand{\)}{\right)}

\def\be{\begin{equation}}
\def\ee{\end{equation}}
\def\ba{\begin{eqnarray}}
\def\ea{\end{eqnarray}}
\def\nn{\nonumber}

\def\p{\partial}
\def\d{\mathrm{d}}
\def\mn{_{\mu \nu}}

\def\({\left(}
\def\){\right)}
\def\mpl{M_{\rm Pl}}

\title{A cautionary case of casual causality}

\author[a]{Calvin Y.-R. Chen,}
\author[a,b]{Claudia de Rham,}
\author[a]{Aoibheann Margalit}
\author[a,b]{and Andrew J. Tolley}

\affiliation[a]{Theoretical Physics, Blackett Laboratory, Imperial College, London, SW7 2AZ, UK}
\affiliation[b]{CERCA, Department of Physics, Case Western Reserve University, 10900 Euclid Ave, Cleveland,
OH 44106, USA}

\emailAdd{calvin.chen16@imperial.ac.uk}
\emailAdd{c.de-rham@imperial.ac.uk}
\emailAdd{a.margalit19@imperial.ac.uk}
\emailAdd{a.tolley@imperial.ac.uk}

\abstract{
We distinguish between the notions of asymptotic causality and infrared causality for gravitational effective field theories, and show that the latter gives constraints consistent with gravitational positivity bounds. We re-explore the scattering of gravitational waves in a spherically symmetric background in the EFT of gravity in $D\ge 5$, for which the leading-order correction to Einstein gravity is determined by the Gauss-Bonnet operator. We reproduce the known result that the truncated effective theory exhibits apparent time advances relative to the background geometry for specific polarisations, which naively signal a violation of causality.
We show that by properly identifying the regime of validity of the effective theory, the apparent time advance can be shown to be unresolvable. To illustrate this, we identify specific higher-dimension operators in the EFT expansion which become large for potentially resolvable time advances, rendering the EFT expansion invalid.
Our results demonstrate how staying within the confines of the EFT, neither infrared nor asymptotic causality are ever violated for Einstein-Gauss-Bonnet gravity, no matter how low the scale, and furthermore its causality can be understood without appealing to a precise UV completion such as string theory.
}

\begin{document}

\maketitle

\setlength{\parskip}{1em}                   


\section{Introduction}\label{sec: intro}
Relativistic causality is a powerful tool in discriminating between low-energy field theories. There is a long history of using causality properties to constrain quantum mechanics and field theory dating back to the 1950's. Two clear approaches emerge\footnote{Another perspective is that causality is restored at the quantum level by means of a Chronology Protection Mechanism even if not manifest at the classical level, \cite{Hawking:1991nk,Burrage:2011cr}. We shall not consider this approach in this work.} -- one is to impose a bound on the scattering time delay  \cite{Eisenbud:1948paa,Wigner:1955zz,Smith:1960zza,Martin:1976iw,de2002time}, which follows from assuming a scattered wave cannot emerge from the scattering region before it arrives. The second is to demand analyticity for scattering amplitudes via dispersion relations \cite{Nussenzveig:1972tcd}. In recent years the former criterion has become known as asymptotic (sub)luminality \cite{Camanho:2014apa,Camanho:2016opx,Goon:2016une,Hinterbichler:2017qcl,Hinterbichler:2017qyt,AccettulliHuber:2020oou}, and the latter has developed into a multitude of positivity bounds which can be used to put powerful constraints on consistent low-energy effective field theories (EFTs) \cite{Pham:1985cr,Ananthanarayan:1994hf,Adams:2006sv,
Arkani-Hamed:2020blm,Chiang:2021ziz,deRham:2017avq,deRham:2017zjm,
Bellazzini:2020cot,Tolley:2020gtv,Caron-Huot:2020cmc,Sinha:2020win,
Du:2021byy,Haldar:2021rri,Raman:2021pkf}.

Both of these approaches are notoriously more subtle when applied to dynamical gravitational theories. Low-energy effective theories can exhibit multiple effective lightcones for different propagating species, even in theories with second-order equations of motion \cite{Babichev:2007dw,deRham:2014zqa,Reall:2021a}, and the metric itself is no longer invariant under field redefinitions, leading to an inherent ambiguity in the local meaning of (sub)luminality. For these reasons it is natural to focus attention on the S-matrix as the main observable. Due to IR divergences in $D=4$, the S-matrix is most clearly defined in $D\ge 5$, and in these dimensions the leading curvature correction to the gravitational effective theory is the Gauss-Bonnet (GB) term.

On asymptotically flat spacetimes, it is straightforward to define a generalised Eisenbud-Wigner time delay $\Delta T$ \cite{Eisenbud:1948paa,Wigner:1955zz,Smith:1960zza,Martin:1976iw} directly from the S-matrix, for the scattering of particles of arbitrary spin, including the graviton.
One plausible definition of relativistic causality is to demand that this total time delay is positive, i.e. there can be no net time advance
\be
\label{eq:asymp net}
\Delta T^{\rm net}> 0 \, .
\ee
This is the condition for the absence of asymptotic superluminalities, and we shall refer to this henceforth as the ``{\it asymptotic causality}" condition. This is the perspective taken in  \cite{Gao:2000ga,Camanho:2014apa}. More precisely, asymptotic causality is really the statement that there should be no resolvable net time advance
\be\label{asympcausality}
\Delta T^{\rm net}\gtrsim -  \omega^{-1} \, ,
\ee
where $\omega$ is the frequency of the scattered wave\footnote{
When a single black hole (BH) or shock wave does not generate a resolvable delay, attempts to generate resolvable time delays by aligning multiple ($N$) configurations
have been proposed in the literature \cite{Adams:2006sv,Camanho:2014apa}.
 We shall consider such setups in a companion paper \cite{Shockwavepaper} and demonstrate that accumulating multiple configurations cannot lead to a resolvable time advance in the regime of validity of the EFT, consistent with the spherically symmetric solutions considered here, and the cosmological solutions considered in \cite{deRham:2020a}. Additional issues with the shock wave arguments have been highlighted in \cite{Papallo2015:gra,Hollowood:2015elj}.}. The resolvability condition is just a reflection of the uncertainty principle and the wave nature of the scattered states. It is important to stress that the asymptotic causality condition is weak in the sense that it allows for propagation of matter and gravitational waves (GWs) that is superluminal with respect to the metric. The reason for this is that already in General Relativity (GR) with no EFT corrections, there can be a non-zero positive time delay $\Delta T^{\rm GR}$ relative to the asymptotic Minkowski metric, known as the Shapiro time delay. Since all matter couples to the metric, the net time delay inferred from any scattering amplitude for any species includes the GR time delay with corrections from interactions which, at low energies, are captured by higher-dimension operators in the EFT
\be\label{eq: time delay split}
\Delta T^{\rm net}= \Delta T^{\rm GR} + \Delta T^{\rm EFT} \, .
\ee
Thus it is possible that matter (including light) or even GWs may travel superluminally with respect to the metric: $\Delta T^{\rm EFT}<0$, and nevertheless lead to a positive net time delay $\Delta T^{\rm net}>0$.
In a gravitational effective theory controlled by a cutoff scale $\Lambda$, we may define
\be\label{eq: GRpart}
\Delta T^{\rm GR} = \lim_{\Lambda \rightarrow \infty} \Delta T^{\rm net} \, ,
\ee
so that $\Delta T^{\rm EFT} $ essentially contains only those terms in the net time delay which come from an inverse expansion in heavy masses. This expansion is only meaningful at low energies, but will be the natural form given within the low-energy EFT.

A more refined notion of causality, still determined by the asymptotic structure of the spacetime, is that the time delay should not be smaller than its value in GR, or more precisely the value obtained in the limit $\Lambda \rightarrow \infty$ when the effects of heavy modes are removed.
This reflects the fact that causality in the local field theory is fixed by the background geometry in which the fields fluctuate, and not the asymptotic Minkowski geometry.
From a field theory perspective, we know that it is the front (and not phase nor group) velocity of propagating modes that defines causality via the support of the retarded propagators. This front velocity is the phase velocity of the high-frequency modes.
The equivalence principle tells us that the high-frequency modes only care about the Minkowski metric of the local inertial frame, and cannot know about the asymptotic metric. The EFT contribution to the time delay $\Delta T^{\rm EFT} $ encodes the effect of interactions of the scattered states with other massive particles, be they heavier standard model particles or higher-spin string states. If $\Delta T^{\rm EFT} $ is allowed to be negative, it still implies that the scattering of the light fields is violating relativistic causality locally, via their interactions with heavy fields. With this in mind we define the notion of ``{\it Infrared Causality}":
\be\label{IRcausality}
\Delta T^{\rm EFT} \gtrsim -  \omega^{-1} \, .
\ee
Since in general $\Delta T^{\rm GR} \ge 0$ in $D\ge 5$, it is apparent the requirement of infrared causality is stronger than that of asymptotic causality. The moniker ``infrared" reflects the fact that this is the pragmatic definition of causality from the perspective of a low-energy (IR) observer.

A strong argument that \eqref{IRcausality} is the appropriate causality condition is obtained by taking a decoupling limit $\mpl \rightarrow \infty$ around a fixed background metric. In this limit, the graviton decouples and the gravitational theory reduces to a field theory on a fixed background metric. If the matter is minimally coupled to the metric at high energies, then the causal structure encoded in the support of the retarded correlation functions will be determined by the spacetime metric. In 1980, Drummond and Hathrell demonstrated that within the low-energy description of quantum electrodynamics (QED), after the electron has been integrated out to one-loop, the photon can appear to propagate outside the lightcone of the background metric on certain curved spacetimes \cite{Drummond:1979pp}. A careful treatment of the validity of the effective theory shows that any resulting time advance is not resolvable \cite{Hollowood:2015elj,deRham:2020a} and so \eqref{IRcausality} is satisfied\footnote{The resolution offered at the time was that the effect is too small to be observed within the resolving power of the EFT: the cumulative distance advance for a photon traversing a BH spacetime, say, is smaller than the Compton wavelength of the electron. This idea was revisited in \cite{Goon:2016une} using a different criteria and also applied to the flat-space Galileon. The perspective here and in  \cite{deRham:2020a,Hollowood:2015elj} is rather that the condition for resolvability is \eqref{IRcausality}.}.
In addition, it was shown by Hollowood and Shore \cite{Hollowood:2007kt,Hollowood:2015elj,Hollowood:2016ryc}, that the issue vanishes in the high frequency limit when microscopic degrees of freedom are brought back to life  (the electron in the case of QED) and that causality is encoded in the background metric. Since QED itself is a consistent, causal theory, even on a gravitational background, it is not surprising that the apparent acausality that appears in its low-energy EFT is only an artifact of the truncation in the macroscopic picture.

Although a given UV completion, such as string theory \cite{DAppollonio:2015fly}, or QED (as a partial UV completion of Euler-Heisenberg theory) may make causality manifest at high energies, it is still important to establish how causality is preserved from the low-energy perspective.
From a pragmatic point of view, this is because we experience Nature through the eyes of low-energy EFTs and rarely have access to the ultimate UV completion to identify causality \cite{Junpaper}.
However, beyond the pragmatic applications, from a conceptual point of view, the main reason we should aim at resolving causality directly within the low-energy EFT is that ultimately any  superluminal propagation, no matter how close to luminal,  can always in principle lead to non-zero correlation functions outside the lightcone at arbitrarily large distances. Since the EFT is supposed to govern the IR, large distance physics, we must be able to diagnose the presence or absence of acausality \emph{entirely within the context of the EFT}, without appealing to a given UV completion.

With this in mind, in \cite{deRham:2020a} it was argued that a low-energy theory corresponding to a consistent UV theory self-protects against apparent causality violations.
In particular, apparent time advances $\Delta T^{\rm EFT}<0$ are never resolvable within the regime of validity of the consistent EFT, and therefore cannot be manipulated to lead to a macroscopically larger lightcone.  Examples considered are the propagation of GWs in the leading-order EFTs describing (a) FLRW and (b) 4-dimensional Schwarzschild spacetimes. It had been shown in previous work \cite{deRham:2019ctd,deRham:2020ejn} that superluminal speeds were possible in both scenarios for a particular sign choice of Wilsonian coefficients in their respective actions. A crucial ingredient in the arguments of \cite{deRham:2020a} is that consistency of the EFT imposes a maximum on the frequency of the scattered wave,
 and this, in turn, keeps the time delay $\Delta T^{\text{EFT}}$ below the resolution scale while remaining in the EFT.

In the present work, we extend the results of \cite{deRham:2020a} by considering a higher-dimensional Schwarzschild BH, with the GB term as the leading correction in the EFT. In agreement with previous works, we will find that regardless of the choice of sign for the Wilsonian coefficient in our EFT, i.e. the sign of the GB term, there are always gravitational degrees of freedom propagating at superluminal speeds\footnote{Specifically, the angular speed receives the correction, while the radial speed remains luminal.} owing to the GB correction \cite{Reall:2014}. This manifests as a time advance $\Delta T^{\rm EFT} < 0$ for those modes relative to the background metric. Taken at face value, this suggests that Einstein-GB gravity violates the ``infrared causality" condition (if the right-hand side of \eqref{IRcausality} were set to vanish). The goal of the present work is to show that this is not the case, provided we understand this correctly as a gravitational effective field theory. Imposing all of the requirements for the validity of the EFT expansion, we will show that the would-be time advance is not resolvable, i.e. the infrared causality condition \eqref{IRcausality} is satisfied for all modes within the regime of validity. Crucially, in order to recognise this, we do not need to appeal to the precise form of the UV completion, be it an infinite number of higher spins, or loops of massive particles. This is also true regardless of how low the scale $\Lambda$ is, and there is no need to tie the scale $\Lambda$ at which the GB terms enter to the string scale or Planck scale. Furthermore since the stronger condition \eqref{IRcausality} is satisfied, the weaker condition \eqref{asympcausality} is automatically satisfied.

To make this clear, we identify a specific dimension-8 curvature operator $\mathcal{L}_{\text{D8}}$, that will generically arise in the EFT expansion, whose contribution to the time delay can be seen to dominate the GB term at high energies. The presence of this and similar higher-dimension operators imposes a cutoff on the scattering energy $\omega$ for which the time delay calculation may be trusted. Imposing this EFT bound on $\Delta T^{\rm EFT}$ shows that it is \emph{unresolvable} and hence respects ``infrared causality" \eqref{IRcausality}. Although we have chosen a specific higher-dimension operator to demonstrate this, it is easy to argue in general terms that there will always be some operators in the EFT that will effectively impose this bound.

To further demonstrate that the condition \eqref{IRcausality} is the correct one, we apply the same arguments to a scalar theory known to violate causality via positivity bound arguments. By considering the scalar theory in a fixed Minkowski background (for which $\Delta T^{\rm GR}=0$), and choosing an analogous spherically symmetric background for the scalar field, we can engineer a situation that closely parallels  the EFT of gravity.
In this case, we find that imposing the bounds implied by the validity of the EFT, it remains possible to engineer a resolvable time advance. Thus the condition \eqref{IRcausality} correctly identifies the acausality of this scalar theory.
We then proceed to consider a scalar Goldstone model with $c(\partial \phi)^4$ interactions that is taken as the poster child example of positivity bounds. We demonstrate that the infrared causality condition \eqref{IRcausality} correctly reproduces the gravitational positivity bounds  conjectured in \cite{Alberte:2020jsk,Alberte:2020bdz} and inferred from  impact parameter bounds in \cite{Caron-Huot:2021rmr}, namely (up to order unity factors),
\be
c \gtrsim - \frac{\Lambda^{D-2}}{\mpl^{D-2}} \, .
\ee
Closely related bounds which utilise additional assumptions\footnote{The stronger bounds considered in \cite{Hamada:2018dde,Tokuda:2020mlf,Herrero-Valea:2020wxz,Alberte:2021dnj} require some knowledge or assumptions about the UV completion. We will not be able to connect with them through our analysis here which is entirely within the low-energy EFT.} in the dispersion relation are discussed in \cite{Hamada:2018dde,Tokuda:2020mlf,Herrero-Valea:2020wxz,Alberte:2021dnj}.

The rest of this paper is organised as follows.
In section \ref{sec: BH in EFT}, we introduce the gravitational low-energy EFT in $D$ dimensions and clarify the scale at which various operators enter. We then  describe  the background Schwarzschild-like solution, consider metric perturbations around it and  provide their governing wave equations.
More details about their parameterisation and effective potentials can be found in appendices \ref{app: master variables} and \ref{app: potentials}.
We proceed by discussing the different notions of causality and derive expressions for the time delays experienced by GWs travelling through the BH spacetime.
As promised, we will see that some polarisations actually experience a time advance due to the first-order correction in the EFT.
In section \ref{sec: validity}, we demonstrate how to deduce the regime of validity for the EFT of gravity and show that the time advance cannot be resolved within the confines of the EFT and hence does not constitute a violation of causality.
In section \ref{sec: galileon toy model}, we repeat the exercise for a scalar field theory in flat spacetime and show that, by contrast, one can obtain a resolvable time advance in this set-up. In section \ref{sec: positivity} we show that, applied to the Goldstone model, the infrared causality bound correctly reproduces the gravitational positivity bounds.
We summarise in section \ref{sec: discussion}, and briefly discuss ongoing work on this topic.

We work in units where $\hbar = c = 1$ and in the mostly-plus signature $(-,+,\dots,+)$.
As we will work in coordinates which make the spherical symmetry of the background manifest, it is useful to introduce different index conventions for the 2-dimensional orbit space (i.e. the $(t,r)$-coordinates)  and the $(D-2)$-dimensional base space (i.e. the coordinates on the sphere $S^{D-2}$).
To this end, tensors on the full $D$-dimensional manifold are indexed with letters from the Greek alphabet.
Tensors on the $(D-2)$-dimensional submanifold $S^{D-2}$ are indexed with letters from the middle of the Latin alphabet $(i, j, \dots)$.
The remaining $(t,r)$ indices are indicated by letters at the beginning of the Latin alphabet $(a, b, \dots)$.


\section{Gravitational effective field theories}
\label{sec: BH in EFT}

We shall be concerned with effective theories of gravity in dimensions $D\ge 5$. For simplicity we focus on the graviton as the only degree of freedom in the low-energy EFT. As such, the low-energy effective theory may be taken to be symbolically of the form
\ba
\label{eq: generic EFT action}
    S_{\rm EFT} &=& \int \d^Dx \sqrt{-g} \, \Mpl^{D-2}\left(\frac{1}{2} R + \Lambda^2\sum_{m\ge 0,n\ge 2} c_{mn} \left(\frac{\nabla}{\Lambda}\right)^m \left(\frac{\text{Riemann}}{\Lambda^2}\right)^n\right) \nonumber \\
    &+& \int \d^D x \sqrt{-g}  \, \tilde \Lambda^{D}\sum_{m\ge 0,n\ge 2} d_{mn} \left(\frac{\nabla}{\tilde \Lambda}\right)^m \left(\frac{\text{Riemann}}{\tilde \Lambda^2}\right)^n  \, .
\ea
There is some redundancy in this parameterisation, which we have introduced to reflect the two main types of contributions. The first line indicates the typical form of corrections that arise from tree-level effects of higher-spin ($s \ge 2$) states of mass $\Lambda$. This is, for example, the form typical of weakly coupled string theories. The second line indicates the typical form of corrections from loops of heavy fields of mass $\tilde \Lambda$, including those of spin $s<2$ (this completion can still be weakly coupled). The latter effects are suppressed parametrically by $(\Lambda/\Mpl)^{D-2}$ relative to the tree-level effects for $\Lambda \sim \tilde \Lambda$, so this acts as the loop counting parameter. From the low-energy point of view, it is in general impossible to know whether a particular higher-curvature term comes from tree-level higher-spin effects or loop contributions and so we must allow for both generic countings. Regardless of the choice of parameterisation, the cutoff of the EFT will be associated with the scale controlling the asymptotic expansion, which is determined by the terms at high $m$ and $n$, for which the distinction in parameterisation is increasingly unimportant. Thus both $\Lambda$ and $\tilde \Lambda$ may be regarded as cutoffs for the low-energy effective theory, indicating the scale at which new physics needs to be introduced to provide a consistent UV completion. 

In what follows we shall consider only the leading-order terms in the EFT of gravity in $D \ge 5$ dimensions, described by the action
\begin{equation}
\label{eq: effective action}
    S_{\text{eff}} = \int \d^Dx \sqrt{-g} \, \Mpl^{D-2}\left(\frac{1}{2} R + \frac{\cGB}{\Lambda^2} R_{\text{GB}}^2 + \dots\right),
\end{equation}
where $R_{\text{GB}}^2 = R_{\mu \nu \alpha \beta}^2 - 4 R_{\mu \nu}^2 + R^2$ is the GB term, $\cGB$ is an $\mathcal{O}(1)$ dimensionless coefficient and $\Lambda \lesssim \Mpl$ is the cutoff scale. Positivity bounds for this truncated theory have been considered in \cite{Bellazzini:2015cra,Cheung:2016wjt}.
While other dimension-4 curvature operators (e.g. $R_{\mu \nu}^2$ or $R^2$) may generically enter at the same order in the EFT expansion, we are specifically interested in vacuum solutions with $R_{\mu \nu} = 0$ to leading order, hence the effect of these operators on the propagation of GWs will be suppressed relative to the effect of the GB term.

For ease of notation, we introduce a small dimensionless parameter $\mu = 1/(\Lambda r_g)^2$, where $r_g$ represents the Schwarzschild radius of the BH in GR.
Throughout this section, we will work only up to linear order in $\mu$ (i.e. leading-order in powers of the inverse cutoff scale, $\Lambda^{-2}$).
Truncation at this order is a reflection of the fact that \eqref{eq: effective action} represents only the first terms in an infinite series of effective operators built from scalar contractions of Riemann tensors and their covariant derivatives.
For now, we are implicitly assuming it is safe to neglect these higher-dimension curvature operators because they would come suppressed by more powers of $\Lambda^{-2}$ compared to the GB term.
In fact, it is exactly this assumption which will define for us the ``regime of validity" in section \ref{sec: validity}.
In the meantime, we just note that calculations to higher order in $\mu$ would be meaningless since we would generically expect corrections from other operators at the same higher orders.

\subsection{Black holes in $D$-dimensional EFT}
\label{subsec: background}
The vacuum Einstein-Gauss-Bonnet equations are
\begin{equation}
\label{eq: EGB eqs}
   G_{\mu \nu} + \frac{2 \cGB}{\Lambda^2} B_{\mu \nu} = 0,
\end{equation}
where
\begin{equation}
\label{eq: GB tensor B}
    B_{\mu \nu} =  4 R_{\alpha \mu \nu \beta} R^{\alpha \beta} + 2 R\indices{_{\mu}^{\alpha \beta \sigma}}R_{\nu \alpha \beta \sigma}- 4 R_{\mu \alpha}R\indices{_{\nu}^{\alpha}}+ 2 R R_{\mu \nu} -\frac{1}{2}  \RGB^2\, g_{\mu \nu} .
\end{equation}
Since the GB tensor $B_{\mu \nu}$ is already suppressed by $\Lambda^{-2}$ compared to the Einstein tensor and we consider a Ricci-flat vacuum solution at leading order, $R_{\mu \nu} = \mathcal{O}(\Lambda^{-2})$, we can ignore any terms in \eqref{eq: GB tensor B} containing a Ricci tensor/scalar without compromising the first-order result. The leading-order static, spherically symmetric and asymptotically flat solution to \eqref{eq: EGB eqs} is
\begin{equation}
\label{eq: background metric}
    \d s^2 = -f(r)\d t^2 + \frac{1}{f(r)}\d r^2 + r^2 \d\Omega_{D-2}^2,
\end{equation}
where $\d\Omega_{D-2}^{2}$ is the line element on the $(D-2)$-sphere $S^{D-2}$, and the metric function is
\begin{equation}
\label{eq: metric function}
    f(r) = 1 - \rgr^{D-3} + 2(D-3)(D-4) \cGB \mu \rgr^{2D-4} + \mathcal{O}(\mu^2)\,.
\end{equation}
The resulting solution is a static Schwarzschild-like BH, with horizon $r_H$ set by $f(r_H)=0$, which differs ever so slightly from the GR Schwarzschild radius $r_g$, $r_H=r_g(1+\mathcal{O}(\mu))$.

\subsection{Metric perturbations}
\label{subsec: metric perturbations}
We are interested in the dynamics of linearised metric perturbations, denoted by $h_{\mu \nu}$, to this curved background.
As we will see, due to the EFT corrections, these modes can follow geodesics that deviate slightly from the null ones, becoming either time-like or space-like,  and thus lead to concerns about causality.

Given the spherical symmetry of the background spacetime, it is useful to parameterise the components of $h_{\mu \nu}$ according to their transformation properties under $SO(D-1)$.
In appendix \ref{app: master variables}, we follow the procedure outlined in \cite{Kodama:2000bra,Kodama:2003a,Ishibashi:2003ap} to reduce the $D(D+1)/2$ components of the symmetric $h_{\mu \nu}$ tensor down to $D(D-3)/2$ scalars called ``master variables".
These master variables entirely encode all the propagating degrees of freedom of the massless spin-2 field associated with metric perturbations.
They fall into three categories --- scalar, vector or tensor --- depending on the $SO(D-1)$-transformation properties of the $h_{\mu \nu}$-components from which they derive.
The vector modes are the higher-dimensional analogue of the Regge-Wheeler axial mode in 4-dimensional GR, and the scalar modes correspond to the Zerilli polar mode in 4-dimensional GR \cite{Regge:1957cc,Zerilli:1970ba}.

The evolution equations for the master variables, known as the ``master equations", are derived from the first-order perturbation to the Einstein-Gauss-Bonnet equations \eqref{eq: EGB eqs}.
Since the GB operator famously produces second-derivative equations of motion, they can be cast in the form of a Schr\"odinger-like wave equation with a potential $V$ sourced by the background curvature:
\begin{equation}
\label{eq: master eq}
    -\frac{\partial^2}{\partial t^2}\Phi_M + f\frac{\partial}{\partial r}\left(f\frac{\partial}{\partial r}\Phi_M\right) - V_M \Phi_M = 0,
\end{equation}
where $M \in \{S,V,T\}$ labels the modes.
As implied by \eqref{eq: master eq}, the potential experienced by the master variables depends only on whether they are classed as scalars ($S$), vectors ($V$) or tensors ($T$).
Each partial wave evolves independently because of the spherical symmetry of the background.
Mode indices are suppressed, but the potentials $V_M$ carry the dependence on the partial wave number $\ell$, as in \eqref{eq: tensor potential} below.
The exact potentials for metric perturbations to static BHs are calculated for full (non-perturbative) Einstein-Gauss-Bonnet gravity in \cite{Dotti:2005a,Dotti:2005b} and for general Lovelock theories in \cite{Takahashi:2009a,Takahashi:2010aa}.
We are only interested in their leading-order behaviour in the EFT expansion, for reasons already discussed above subsection \ref{subsec: background}.
All three potentials are provided explicitly to leading order in $\mu$ in Appendix~\ref{app: potentials}.
The dynamics of the tensor modes, specifically, will become of particular interest in subsection \ref{subsec: validity from time delays} so we reproduce its potential here, for reference:
\begin{align}
\label{eq: tensor potential}
    \frac{V_{T}}{f} = &\frac{1}{r^2}\bigg[k_{T}^2\left(1 + 8 \cGB \mu (D-1) \rgr^{D-1}\right) \nonumber \\
    &+ \frac{D(D-6)+16}{4}\left(1 - 32\cGB \mu \frac{(D-1)(D-6)}{D(D-6)+16} \rgr^{D-1}\right) \\
    &+ \frac{(D-2)^{2}}{4}\rgr^{D-3}\left(1 - 2\cGB \mu \frac{(D-4)\left[3D(D-3)(D-6)-32\right]}{(D-2)^{2}}\rgr^{D-1}\right)\bigg]. \nonumber
\end{align}
Here, $-k_T^2$ is the eigenvalue of the Laplace-Beltrami operator on the $(D-2)$-sphere acting on a tensor-type spherical harmonic, as discussed further in appendix \ref{subapp: tensor modes}.
It is related to the more familiar integer mode numbers $\ell$ by $k_T^2 = \ell(\ell+D-3)-2$, where $\ell=1,2,\ldots$.
The expression for $V_T$ is by far the simplest of the three potentials.


\subsection{Apparent local superluminality}
\label{sec: apparent acausality}
Before proceeding to the calculation of the scattering time delay, we can already get a local indication of the supposed acausality by inspecting the expression for the tensor potential \eqref{eq: tensor potential}.
Specifically, it is apparent that the angular speed $v_{\Omega}$ of the GWs is modified by the GB contribution.
(Notably, the radial speed is unaffected to leading order in the EFT.)
By identifying $v_{\Omega}^2$ with the coefficient of $k_T^2/r^2$, we immediately see that
\begin{equation}
    v_{\Omega}^2 = 1 + 8 \cGB \mu (D-1) \rgr^{D-1}.
\end{equation}
If $\cGB$ were a positive number, the angular speed of the GWs would be locally superluminal ($v_{\Omega}>1$).
In principle, such superluminality could lead to a serious violation of causality.
While all other massless particles, of any frequency, are confined to the lightcone set up by the background metric, low-frequency GWs could propagate \textit{slightly outside} that lightcone. A number of works have considered the implications of this apparently enlarged lightcone \cite{Burrage:2011cr,Reall:2014,Benakli:2015qlh,Papallo2015:gra,Andrade:2016yzc,Brustein:2017iet,Sherf:2018uth,Caceres:2019pok}.

At this point, one might be tempted to forbid $\cGB > 0$ and content oneself that any causal UV complete theory should have $\cGB \leq 0$ as part of its low-energy description.
Unfortunately, examination of the vector potential \eqref{eq: vector potential app} suggests that those modes would propagate superluminally if $\cGB < 0$.
Thus, it seems we should be forced to conclude that $\cGB = 0$ is the only viable option and that the GB term is not a good causal operator.

This argument is altogether too quick for two reasons. First, local superluminalities do not by themselves necessarily indicate causality violation. What is in tension with causality is the possibility of creating closed time-like curves which requires building up superluminalities over some trajectories, which  is why we will compute the scattering time delay below. Within the time delay,  local superluminalities at one point could be compensated by subluminalities at other points. The second reason why local superluminalities cannot by themselves indicate whether causality is necessarily violated is because we have yet to impose the full requirements for the validity of the EFT which, as we shall see, will mitigate the level of superluminality one can ever enjoy.

\subsection{Scattering phase shifts}
\label{subsec: phase shifts}

To calculate the scattering time delay \eqref{eq: EW time delay}, it is convenient to first calculate the phase shift $\delta_{\ell}$ of a partial wave scattering to/from asymptotic infinity in the BH spacetime. Since the master equations \eqref{eq: master eq} are reminiscent of the Schr\"odinger equation, we may utilise the semiclassical Wentzel-Kramers-Brillouin (WKB) approximation to determine the phase shift. This will automatically reproduce the classical time delay and the eikonal (shock wave) time delay \cite{AccettulliHuber:2020oou} in appropriate limits, as discussed in the appendices of \cite{deRham:2020a}.

It is usual to consider the BH perturbation equations in the tortoise coordinate $\hat{r}$ defined via
\begin{equation}
\label{eq: tortoise coord}
    \d r=f(r)\d\hat{r} \, ,
\end{equation}
however, in order to compare with standard scattering problems (without a horizon) we rather work with Langer coordinates $r = e^{\rho}$. In the case of a regular source such as a star, this has the effect of mapping the origin of coordinates, $r=0$, to $\rho=-\infty$, so that a WKB approximation does well near $\rho=-\infty$ as well as $\rho = \infty$. This is well known to improve the form of the WKB approximation at small $\ell$. Despite the presence of a horizon, this proves to be useful here. We will, however, only be interested in the solution outside the horizon,  $r>r_H$.
Making a wave ansatz for the time-dependence of the master variable $\Phi = e^{-i\omega t} \phi(\rho)$, and defining $\phi= (f e^{-\rho})^{-1/2} \chi$, the master equation \eqref{eq: master eq} then takes the form
\begin{equation}
\label{eq: master eq tortoise}
    \frac{\d^{2} \chi}{\d\rho^{2}}+ W\left(\rho\right) \chi = 0\,,
\end{equation}
where
\begin{equation}
    W(\rho) = \( \omega^{2}-V(\rho) \) \frac{e^{2 \rho}}{f^2}- \frac{1}{2}\Omega''(\rho)-\frac{1}{4} (1-\Omega'(\rho))^2\,,
\end{equation}
and $\Omega = \ln (f)$. The field $\chi$ should carry both the index $M$ to label its potential and, additionally, a mode number $\ell$.
Each partial wave $\ell$ for each perturbation type $M$ evolves independently according to \eqref{eq: master eq tortoise}.

A property of the potentials $V_{M}$ (as given in appendix \ref{app: potentials}) is that all three agree in the limit $r_g\rightarrow 0$.
This allows us to define a universal impact parameter $b$ via
\begin{equation}
\label{eq: impact parameter}
    b^2\omega^2 = \lim_{r_g\rightarrow 0} \left(\omega^2 e^{2\rho}- W(\rho)\right)\,,
\end{equation}
which results in
\be
b=\omega^{-1} (\ell+(D-3)/2))\,.
\ee
The point of closest approach, or turning point, $\rho_t$, is determined by $W(\rho_t) = 0$.
The turning point coincides with the impact parameter $b$ when $r_g=0$, but the curvature-corrections will depend on whether we are considering scalar-, vector- or tensor-type perturbations.

The master equation in \eqref{eq: master eq tortoise} is now in a suitable form to use the WKB approximation. Although the effective potential goes to infinity at the horizon, what is important for us is the potential barrier around $r \sim b$. For $\rho<\rho_{t}$, we have a barrier with $W<0$, so this is the classically forbidden region. Provided the barrier is wide enough for the chosen frequencies, the solution inside the barrier will be dominated by the exponentially decaying solution %
\begin{equation}
	\chi(\rho)\approx \frac{\bar{\chi}}{\left(-W\right)^{1/4}} \exp{\left(-\int_{\rho}^{\rho_{t}} \sqrt{-W} \d \rho \right)},
\end{equation}
where $\bar{\chi}$ is a constant. Using the WKB connection formula, we find that for $\rho>\rho_{t}$
\begin{equation}
\label{eq: wave solution}
	\chi(\rho)\approx \frac{2\bar{\chi}}{W^{1/4}} \sin{\left(\int_{\rho_t}^{\rho}\sqrt{W}\d \rho+\frac{\pi}{4}\right)}.
\end{equation}
We define the scattering phase shift $\delta_{\ell}$ by demanding that this solution has the asymptotic form
\begin{equation}
\label{eq: partial wave asymptotic}
    \chi_{\ell} \stackrel{\rho \rightarrow \infty}{\sim} \left(e^{2 i \delta_{\ell}} e^{i\omega r}+e^{i \pi \ell} e^{i \pi (D-2)/2} e^{-i\omega r}\right).
\end{equation}
We have re-introduced the partial wave label $\ell$ to emphasise that the phase shift is an $\ell$-dependent quantity.
Comparing the true solution \eqref{eq: wave solution} with the asymptotic form \eqref{eq: partial wave asymptotic}, we find the phase shift in terms of the original radial coordinate
\begin{equation}
	\delta_{\ell}(\omega)=\int_{\rho_{t}}^{\infty}\left(\sqrt{W_{\ell}}-\omega e^{ \rho } \right)\d \rho-\omega r_{t}+\frac{\pi}{2}(\ell+(D-3)/2)) ,
\end{equation}
where $r_{t}=r\left(\rho=\rho_t\right)$. In the limit $r_g \rightarrow 0$ this formula correctly gives $\delta_{\ell}(\omega)=0$.

\subsection{Time delays}
\label{subsec: time delays}
We now define the so-called ``Eisenbud-Wigner time delay" \cite{Eisenbud:1948paa, Wigner:1955zz, Smith:1960zza, Martin:1976iw}. Consider an incident gravitational wave packet for a given partial wave of mode number $\ell$ peaked around a frequency $\omega$ traversing a BH spacetime. The time delay describes the amount by which the scattered wave packet is delayed relative to the same wave packet propagating on Minkowski spacetime. For each partial wave the net time delay is
\begin{equation}
\label{eq: EW time delay}
    \Delta T_{\ell} = 2\left.\frac{\partial\delta_{\ell}(\omega)}{\partial\omega}\right|_{\ell}.
\end{equation}
While this is the time delay most appropriate to think about spherical wave scattering, for plane wave scattering it is often useful to consider the phase shift in the large $\ell$ limit and work at fixed impact parameter $b \sim \omega^{-1} \ell$. This is the limit in which the partial wave expansion becomes a Fourier transform. It is thus common to also work with the time delay at fixed impact parameter
\begin{equation}
\label{eq: EW time delay2}
    \Delta T_{b} = 2\left.\frac{\partial\delta_{\ell}(\omega)}{\partial\omega}\right|_{b}.
\end{equation}
The two time delays are qualitatively similar and are clearly related. It is the latter that is used in \cite{Camanho:2014apa} within the eikonal limit, and reproduces the classical GR time delay. However when considering waves, it is the former which is more appropriate as it is meaningful even for $\ell=0$. The precise distinction between the two will not be important in our following discussion, and both are easily calculated from the scattering phase shift.

Since every field couples to gravity, every species will receive a contribution $\Delta T^{\text{GR}}$ to the time delay from the gravitational background, what is known in $D=4$ as the Shapiro time delay. In $D \ge 5$ the equivalent Shapiro time delay is IR finite and positive for a positive mass source. In addition, scattering particles will receive an additional time delay from interactions with heavy states. The latter effect is what is captured at low energies by the higher-dimension operators in the EFT. Thus it is natural to separate the low-energy time delay in the manner
\begin{equation}
    \Delta T^{\rm net}= \Delta T^{\text{GR}} + \Delta T^{\rm EFT}.
\end{equation}

As discussed in the introduction, there are two separate notions of causality used in the literature:
\begin{enumerate}
    \item {\bf Asymptotic causality}, \cite{Gao:2000ga,Camanho:2014apa}:
    This is the notion that only the sign of the net time delay is relevant.
    It is typically considered that as long as $\Delta T^{\rm net}> 0$, i.e. it is a true time \textit{delay} with respect to the asymptotic Minkowski spacetime, then there is no violation of causality, however the precise statement of asymptotic causality should still include a notion of resolvability. More specifically, asymptotic causality is the requirement that there is no resolvable net time advance $\Delta T^{\rm net}  \gtrsim -\omega^{-1}$.

    \item {\bf Infrared causality}:
    This is the notion that only the sign of the EFT correction $\Delta T^{\rm EFT}$ to the time delay is relevant. This is the definition that is implicit in considering field theories in curved space, as in the investigations of causality in the EFT description of QED \cite{Drummond:1979pp,Hollowood:2007kt,Hollowood:2015elj}. More precisely we require that $\Delta T^{\rm EFT}\gtrsim -\omega^{-1}$.
    \end{enumerate}

The notion of infrared causality is motivated by the fact that $\Delta T^{\text{GR}}$ is universal for all particles coupled to gravity, including gravitons, and should be considered as the ``reference" time delay. For instance, in cases where the EFT corrections come from integrating out loops of massive fields, which are minimally coupled to gravity in the UV, it is always the case that the causal structure is determined by the high-energy modes which are themselves minimally coupled and so lead to $\Delta T^{\text{GR}}$.

Overall a resolvable negative $\Delta T^{\rm EFT}$ (i.e. a violation of infrared causality) would indicate the fact that some species (in this case gravitons) can enjoy a spacelike geodesic which departs from the null geodesic by a resolvable amount. This would also mean that gravitons would experience a resolvable time advance as compared to other massless particles (say photons) that minimally couple to gravity.

One may worry that the definition of the split $ \Delta T^{\rm net}= \Delta T^{\text{GR}} + \Delta T^{\rm EFT}$ is sensitive to field redefinitions of the metric that change the GR part.  This is partly mitigated by our definition of the GR part \eqref{eq: GRpart} which can be defined at the level of the S-matrix. Furthermore, performing a field redefinition in the low-energy EFT also redefines the metric to which the high-energy modes couple, and in this analysis we consider a frame where at least one field couples minimally to gravity in the UV (at scales well above $\Lambda\lesssim \mpl$).

\subsection{Time delay for $D$-dimensional black hole}

Although the WKB expression for the time delay is capable of dealing with small $\ell$, it will be sufficient to focus on large $\ell \sim b \omega \gg 1$, since the high-energy regime will be important in our subsequent analysis.
We will also assume for simplicity that the scattering impact parameter is large compared to the Schwarzschild radius, $b \gg r_g$.
With these assumptions, and using the effective potentials provided in Appendix \ref{app: potentials}, we find the following expressions for the time delays\footnote{These are the time delays at fixed $\ell$. Very similar expressions follow for the time delay at fixed $b$.}
\begin{subequations}
\label{eq: all time delays}
\begin{align}
\label{eq: tensor time delay}
    \Delta T_T &= \Delta T^{\text{GR}} \left[1-\frac{8(D-1)}{D-3}\cGB \mu \left(\frac{r_{g}}{b}\right)^{2}\right], \\
\label{eq: vector time delay}
    \Delta T_V &= \Delta T^{\text{GR}} \left[1+\frac{4(D-1)(D-4)^{2}}{D-3}\cGB \mu \left(\frac{r_{g}}{b}\right)^{2}\right], \\
\label{eq: scalar time delay}
    \Delta T_S &= \Delta T^{\text{GR}} \left[1 + \frac{8(D-1)(D-4)^{2}}{D-3} \cGB \mu \left(\frac{r_{g}}{b}\right)^{2}\right],
\end{align}
\end{subequations}
where
\begin{equation}
\label{eq: GR time delay}
    \Delta T^{\text{GR}}=\frac{(D-2)\sqrt{\pi}}{2}\frac{\Gamma\left(\frac{D-4}{2}\right)}{\Gamma\left(\frac{D-3}{2}\right)}\left(\frac{r_{g}}{b}\right)^{D-3}b \, ,
\end{equation}
is the time delay from GR in the same limits.
Note that with particular choices for $\mu$ and $\cGB$, this reproduces the results in \cite{Camanho:2014apa} and \cite{Reall:2014}.

We are now in a position to see how GB theory could appear to violate both asymptotic causality and infrared causality if the resolvability condition was not properly accounted for in the regime of validity of the EFT.
Looking at the expressions in equations \eqref{eq: all time delays}, we see that the EFT corrections to the scalar and vector mode time delays have opposite sign to the tensor mode time delay.
There are no non-zero values of $\cGB$ which can make $\Delta T^{\rm EFT}>0$ for all three.
This immediately suggests a violation of infrared causality.
To generate an asymptotic acausality, on the other hand, we need to further push $b^2 \rightarrow \mu r_g^2 = \Lambda^{-2}$ and then we would find ourselves in a situation where there would be at least one negative $\Delta T_M^{\rm EFT}$ that could overwhelm $\Delta T^{\text{GR}}$ for any non-zero $\cGB$.
This corresponds to the situation in \cite{Camanho:2014apa} where, in their notation, $b^2 \sim \lambda_{\text{GB}}$. The central claim of what follows is that once we have properly identified the regime of validity of the EFT, neither causality conditions are violated in a resolvable way. To see why, we now turn to the consideration of the regime of validity.


\section{Validity of the gravitational EFT}
\label{sec: validity}
In this section, we explore the constraints on the applicability of the EFT description and thereby determine its regime of validity.
The discussion in this section closely mirrors the analysis in \cite{deRham:2020a}. %
\subsection{Constraints on background}
\label{subsec: constraints on background}
As repeatedly emphasised, our action \eqref{eq: effective action} describes just the lowest-order terms in an infinite series of effective operators generically present in the EFT of gravity.
No real example of low-energy expansions of UV-complete theories \cite{Gross:1986iv, Metsaev:1986yb} would ever truncate after $R^2$--operators. Rather, they generically contain all possible dimension-6, -8, -10,\dots operators built out of the curvature and its derivatives, and we have parameterised the typical form in \eqref{eq: generic EFT action}.
Ultimately any EFT expansion is an expansion in derivatives, and so control over this expansion is lost whenever that curvature or its derivatives become too large.  In practice, this means that the EFT is only a useful description of gravity in regions of spacetime where, schematically,
\begin{equation}
\label{eq: schematic background bound}
    \nabla^{2m} \text{Riemann}^n \ll \Lambda^{2m+2n},
\end{equation}
where $\Lambda$ is the cutoff of the effective theory. What is meant by this schematic expression is: all possible scalar contractions built out of $2m$ covariant derivatives and $n$ powers of Riemann tensors.

We may first apply these constraints to the background geometry for which ``Riemann" may be replaced with the Riemann tensor as calculated in the GR background spacetime, since any corrections to those components will be automatically suppressed by additional powers of $1/\Lambda$ in a region where the expansion is controlled. Since our background is Ricci-flat at leading order, it is sufficient to place these bounds on the Weyl or Riemann tensor. Thus an example requirement for the validity of the EFT expansion is
\be
\Box^m \(R_{\mu\nu\alpha\beta} R^{\mu\nu\alpha\beta} \)^n \ll \Lambda^{2m+4n}.
\ee
For the background geometry, Riemann scales as $r_g^{D-3}/r^{D-1}$, and the strongest bound will be at the point of closest approach, which is related to the impact parameter $b$.
The implication is
\begin{equation}
\label{eq: practical background bound}
    \nabla^{2m} \text{Riemann}^n \sim \frac{r_g^{(D-3)n}}{b^{(D-1)n+2m}} \ll \Lambda^{2m+2n}.
\end{equation}
In the limit $n \rightarrow \infty$ for fixed $m$, this is the requirement that%
\begin{equation}
\label{eq: Vainshtein radius for gravity}
    \frac{r_g^{D-3}}{b^{D-1}} \ll \Lambda^2 \, ,
\end{equation}
which is just our usual statement that the curvature is small. At the other extreme, we can consider operators with a large number of covariant derivatives and take $m \rightarrow \infty$ to obtain:
\begin{equation}
\label{eq: bound on b for gravity}
   b \gg \Lambda^{-1}.
\end{equation}
From this statement alone, it is already clear that there is no violation of asymptotic causality \eqref{eq:asymp net} in this EFT if the background is under control, no matter the scale $\Lambda$.

Taken together, \eqref{eq: Vainshtein radius for gravity} and \eqref{eq: bound on b for gravity} define the range of distance scales $b$ for which the EFT expansion is well-defined on a Schwarzschild background.
They both express the fact that our understanding of the low-energy breaks down at short scales or high energies, where the microscopic degrees of freedom that have been integrated out to lead to this EFT ought to be described in their own right. Note that {\it a priori} we do not require $r_g > \Lambda^{-1}$ since it is sufficient that the geometry is asymptotically Schwarzschild at distances $r,b \gg \Lambda^{-1}$, and we do not need to resolve the horizon.
\subsection{Constraints on perturbations}
\label{subsec: constraints on perturbations}

In addition to the largely familiar constraints on the background, we should also ensure that higher-order operators do not spoil the equations of motion for the perturbations we have considered so far. Naively one may expect that, so long as the background is under control, the higher-order EFT corrections to the linear perturbation equations will also be under control. In practice, for backgrounds that enjoy a high level of symmetry (as is the case for the static and spherically symmetric BH solution), the perturbations are able to probe much more of the information contained in the Riemann tensor and hence can provide a richer insight on the validity of the EFT.

From the point of view of the perturbations, since we are dealing with a low-energy EFT, it is clearly not possible to consider scattered particles of arbitrarily high energy. Not only will the Riemann curvature itself be perturbed, the covariant derivatives in generic higher-derivative operators can now act both on the background and on the perturbations. Focusing on a GW of momentum $k^{\mu}$, and replacing all derivatives acting on the perturbations by their form at high energies, i.e. $\nabla_{\mu} h_{\alpha \beta} = i k_{\mu} h_{\alpha \beta}$, the schematic form of the corrections to the equation of motion for GWs $h_{\mu\nu}$ are
\be
k^2 h + \sum_{n+m+p>1}   \frac{c_{nmp}}{\Lambda^{2(n+m+p-1)}}\(\nabla^{p}  {\text{Riemann}}^m\) k^{2n+p} h =0\,,
\ee
where we understand $\nabla$ and $\text{Riemann}$ to be background quantities. Now, although this is a tensor equation, i.e. of the form
\be
O_{\mu \nu}^{\alpha \beta , \text{GR}} h_{\alpha \beta}+O_{\mu \nu}^{\alpha \beta , {\rm EFT}} h_{\alpha \beta}=0\,,
\ee
in order estimate the size of the EFT corrections, it is sufficient to focus on the eigenvalues of $O_{\mu \nu}^{\alpha \beta , {\rm EFT}}$, which are all scalar quantities, relative to $k^2$. Thus to establish the regime of validity for perturbations it is sufficient to consider scalar operators built in the manner
\begin{equation}
\label{eq: schematic perturbed operators}
\left(\frac{\nabla}{\Lambda}\right)^p \left(\frac{\text{Riemann}}{\Lambda^2}\right)^m \left(\frac{k}{\Lambda}\right)^{2n+p} \ll 1\,,
\end{equation}
and again we consider all possible scalar contractions.
Due to the symmetries of the Riemann tensor, and the fact that $k_{\mu}k^{\mu} \approx 0$ on-shell, the tensor with the highest power of $k$ we can build from a single Riemann tensor is
\begin{equation}
\label{eq: tensor A}
    A\indices{^{\mu}_{\nu}} = R\indices{^{\mu}_{\alpha \nu \beta}}k^{\alpha}k^{\beta}.
\end{equation}
The trace of this object vanishes because of the symmetry of the background spacetime, but scalars can be built out of higher powers of $A\indices{^{\mu}_{\nu}}$.
The requirement that they are not too large:
\begin{equation}
\label{eq: bound on Tr[An]}
    \text{Tr}\left[A^n\right] \ll \Lambda^{4n} \, ,
\end{equation}
translates into a bound on the frequency $\omega$ of the wavevector $k_{\mu}$. Denoting the metric on the $(D-2)$-sphere as $\d \Omega^2_{D-2} = \d \theta^2 + \sin(\theta)^2 \d \Omega^2_{D-3}$, and solving the geodesic equation for a massless particle, we can parameterise the scattering momenta as
\begin{equation}
    k^{\mu} \partial_{\mu}=\frac{\omega}{f} \partial_t + \omega\sqrt{1-\frac{f b^{2}}{r^{2}}} \partial_r + \frac{b\omega}{r^{2}} \partial_{\theta} \, ,
\end{equation}
so that
\begin{align}
    A_{\mu\nu}\d x^{\mu}\d x^{\nu}=&\frac{(D-3)r_{g}^{D-3}\omega^{2}}{r^{D-1}}\bigg[-\frac{1}{2}\left((D-2)-(D-1)\frac{b^{2}}{r^{2}}f\right)\d t^{2}
    -\frac{1}{2f^{2}}\left((D-2)-\frac{b^{2}}{r^{2}}f\right)\d r^{2} \nonumber \\
    &+\frac{(D-2)}{f}\sqrt{1-\frac{fb^{2}}{r^{2}}}
    \d t\, \d r +\left(-\frac{b}{2}\d t+\frac{b}{2f}\sqrt{1-\frac{f b^{2}}{r^{2}}}\d r+\frac{b^{2}}{2}\d\theta\right)\d\theta \nonumber \\
    &+\frac{b^{2}}{D-3} \d \Omega^2_{D-3}\bigg].
\end{align}
Evaluating equation \eqref{eq: bound on Tr[An]} explicitly with $n=2$, we find the following bound:
\begin{equation}
C_D \frac{\omega^2 b^2 r_g^{D-3}}{r^{D+1}} \ll \Lambda^4,
\end{equation}
where the coefficient $C_D$ is a dimensionless order-$1$ dimension-dependent factor, $C_D^2=D^{4}-8D^{3}+23D^{2}-26D+6$.
Again, considering this at the point of closest approach $r \sim b$, this condition becomes essentially
\begin{equation}
\label{eq: frequency upper bound 1}
    \omega^2 \ll \frac{\Lambda^4 b^{D-1}}{r_g^{D-3}}.
\end{equation}
No further constraint is obtained from considering the limit $n \rightarrow \infty$. Equation \eqref{eq: frequency upper bound 1} is our first upper bound on the energy of GWs which can safely propagate within the regime of validity of the EFT.

We find another type of bound on the frequency from operators with covariant derivatives.
Contractions of covariant derivatives with Riemann tensor indices vanish, as do contractions of $k_{\mu}$ with $A_{\mu\nu}$, so amongst the remaining possibilities are
\begin{subequations}
\begin{align}
   &\big(\left(k^{\mu}\nabla_{\mu}\right)^m A^{\alpha\beta}\big) \big(\left(k^{\nu}\nabla_{\nu}\right)^n A_{\alpha\beta}\big) \ll \Lambda^{8+4(m+n)},\\
   &\left(k^{\mu}\nabla_{\mu}\right)^m\left(W^{\alpha \beta \gamma \delta}W_{\alpha \beta \gamma \delta}\right) \ll \Lambda^{4+2m}.
\end{align}
\end{subequations}
In the limit $m, n \rightarrow \infty$, the above bounds amount to $k^{\mu}\nabla_{\mu} \ll \Lambda^2$, or what is tantamount to
\begin{equation}
\label{eq: frequency upper bound 2}
    \omega \ll \Lambda^2 b \, ,
\end{equation}
which is our second upper bound on the energy of GWs in the EFT. As long as we consider scattered waves for which $b \gtrsim r_g$, the bound \eqref{eq: frequency upper bound 2} is always stronger than \eqref{eq: frequency upper bound 1}, and so for all intents and purposes we may regard the maximum frequency for which the effective theory can be trusted to be $\omega \sim \Lambda^2 b$. This will prove important in what follows.

It is important to stress that we are allowed to consider frequencies $\omega$ larger than $\Lambda$, and still remain within the validity of the EFT. The reason for this is that $\omega$ is not a covariant/Lorentz invariant quantity, and in a Lorentz invariant theory bounds should always be placed on Lorentz scalars.  The bound \eqref{eq: frequency upper bound 2} can be intuitively understood as the bound associated with the Mandelstam invariant $k_{\mu} \kappa^{\mu} \ll \Lambda^2$, where $\kappa^{\mu} $ is the typical momentum of the coherent gravitons that make up the background. The bound \eqref{eq: bound on b for gravity} is similarly the requirement that $\kappa_{\mu} \kappa^{\mu} \ll \Lambda^2$.

\subsection{Regime of validity from time delays}
\label{subsec: validity from time delays}
The arguments of the previous section on the regime of validity were largely schematic, and so it is useful to consider explicit examples of higher-dimension operators from which the bounds may be observed.
We will now show how we can arrive at the same frequency bound \eqref{eq: frequency upper bound 1} through concrete calculation with a specific higher-dimension operator. We will concentrate only on tensor-type perturbations to illustrate our point.
Among all the dimension-8 curvature operators expected to enter the EFT \eqref{eq: generic EFT action}, the following one is quite generic
\begin{equation}
\label{eq: S8 extension}
    S_{\text{D8}} = \int \d^{D}x \sqrt{-g}\Mpl^{D-2}\left(\frac{c_{\text{R4}}}{\Lambda^6}R_{\alpha \beta \gamma \delta}R^{\gamma \delta \zeta \iota}R_{\zeta \iota \kappa \lambda}R^{\kappa \lambda \alpha \beta}\right),
\end{equation}
where $c_{\text{R4}} \sim \mathcal{O}(1)$ is our new Wilsonian coefficient. The presence of this operator is very generic to any gravitational EFT and makes no assumption on its precise UV completion.  The truncated  EFT action we now consider is $S_{\text{eff}} + S_{D8}$, where $S_{\text{eff}}$ is given in \eqref{eq: effective action}.  There are of course dimension-6 operators as well as other dimension-8 operators that enter the EFT, however for the purpose of this argument they have no impact as will be discussed later. We will show that by demanding the time delay induced by this particular dimension-8 is subdominant to the GB-induced time delay, a necessary condition for the latter to be taken seriously, we can exactly reproduce \eqref{eq: frequency upper bound 1}.
For clarity of notation, we split the EFT time delay into two components:
\begin{equation}
    \Delta T^{\rm EFT} = \Delta T^{\rm GB} + \Delta T^{\rm D8}+ \dots,
\end{equation}
where $\Delta T^{\rm GB} \propto \mu$ is the GB-correction we have already calculated and $\Delta T^{\rm D8} \propto \mu^3$ will arise from the newcomer.

This new operator will modify the propagation of GWs by introducing higher-derivative, $\mathcal{O}(\mu^3)$-corrections to the master equations \eqref{eq: master eq}.
As usual in an EFT, the appearance of higher derivatives does not signal new ghostly degrees of freedom, and following \cite{deRham:2019ctd} we can trade higher radial derivatives for higher powers of $\omega$ and $\ell$ using the lower-order equations of motion.
The net effect is a correction to the effective potential, which we will denote by $V_T^{\rm D8}$.
Since we are working only to leading order in large-$\ell$ with the WKB approximation,  we only need to know the large-$\ell$ behaviour of $V_T^{\rm D8}$ to calculate the time delay.
This can be deduced by following only the highest-, four-derivative terms, which is done explicitly in appendix \ref{app: higher dim ops}.
The result is
\begin{equation}
\label{eq: V3 potential}
    \frac{V_T^{\rm D8}}{f} = -16(D-1)^{2} c_{\text{R4}} \mu^3 \frac{1}{r^2} \rgr^{2D}  k_T^4 \(1 + \mathcal{O}\left( k_{T}^{-2}\right)\).
\end{equation}
Including these contributions to the tensor modes time delay, we find that the $\mathcal{O}(\mu^3)$-correction is
\begin{equation}
\label{eq: T3 time delay}
    \Delta T_T^{\rm D8} = 16 c_{\text{R4}}\mu^{3}\frac{\sqrt{\pi}(D-1)\Gamma\left(D+\frac{1}{2}\right)}{\Gamma\left(D-1\right)}\left(\frac{r_{g}}{b}\right)^{2D}b^{3}\omega^{2}+ \dots,
\end{equation}
up to terms which are subleading at large $\ell$.
Now, if our original truncated EFT action \eqref{eq: effective action} was to be trusted, it should be the case that $\Delta T_T^{\rm D8}$ is negligible compared to the leading-order EFT term $\Delta T_T^{\rm GB}$ in \eqref{eq: tensor time delay}.
Assuming that the Wilson coefficients are order unity, this is the statement that
\be
\mu^{3} \left(\frac{r_{g}}{b}\right)^{2D}b^{3}\omega^{2} \ll \mu \left(\frac{r_{g}}{b}\right)^{D-1}b\,,
\ee
which, after a little rearrangement gives exactly the upper bound on $\omega$ given in \eqref{eq: frequency upper bound 1}.

Our reason for focusing our attention on \eqref{eq: S8 extension} is that it is the lowest-order term in the generic EFT expansion which introduces higher powers of $\omega^2$ in the effective potential and hence in \eqref{eq: T3 time delay} as compared to \eqref{eq: tensor time delay} (via their conversion to a potential term featuring $k_T^4$).
It is the contrast in frequency dependence of \eqref{eq: T3 time delay} and \eqref{eq: tensor time delay} that leads to the non-trivial bound given in \eqref{eq: frequency upper bound 1}. The fact that no lower-dimension operator gives the same result is a quirk of the symmetries of the background spacetime, combined with the transverse-traceless structure of the tensor spherical harmonics.
This eliminates all other dimension-4 and dimension-6 operators from contention, as explained in appendix \ref{app: higher dim ops}.
In the analogous argument of subsection \ref{subsec: constraints on perturbations}, the dimension-4 scalar $\text{Tr}[A]$ vanished and we had to look to its dimension-8 counterpart $\text{Tr}[A^2]$ to place a non-trivial bound on the wave's momentum.

Of course one could imagine that the EFT happened to have $c_{\text{R4}}=0$ by virtue of symmetry or some sum-rule. However, the special properties of this dimension-8 operator \eqref{eq: S8 extension} that provide the right bound are not unique, as previously mentioned. Any operator that gives terms in the potential depending on $k_T^{2n}$ for $n>1$ (after perturbatively replacing higher derivatives with the lower-order master equation) will do.
This means that, to escape the conclusion that the time delay is unresolvable \eqref{eq: GB time delay is unresolvable}, one would have to set to zero essentially all higher-derivative operators, giving an EFT that could never have come from a consistent UV completion. Indeed it is now well understood from positivity bound arguments that there are upper and lower bounds on generic Wilson coefficients at every order in the derivative expansion \cite{Tolley:2020gtv,Caron-Huot:2020cmc,Bern:2021ppb}. This means that, in general, setting coefficients of higher-curvature operators to zero is not necessarily even an option. This conforms with general expectations from explicit loops from massive fields, where essentially every curvature invariant is generated at some order.

\subsection{Unresolvability of time delay}
\label{subsec: resolvability}
We are now in a position to demonstrate our central result. To take any apparent acausality seriously, it must be resolvable within the confines of the EFT as defined by the four bounds: \eqref{eq: Vainshtein radius for gravity}, \eqref{eq: bound on b for gravity}, \eqref{eq: frequency upper bound 1} and \eqref{eq: frequency upper bound 2}. Focusing on the order of magnitude of the GB contribution we find
\begin{equation}
\label{eq: GB time delay is unresolvable}
    \omega \abs*{\Delta T^{\text{GB}}} \sim \mu \omega b \left(\frac{r_g}{b}\right)^{D-1} \sim \omega b \frac{1}{b^2 \Lambda^2} \left(\frac{r_g}{b}\right)^{D-3} \ll \left(\frac{r_g}{b}\right)^{\frac{D-3}{2}} \ll 1\,,
\end{equation}
where we have used \eqref{eq: frequency upper bound 1} to bound the frequency $\omega$ and the fact that $b \gg r_g$.
In fact, using \eqref{eq: frequency upper bound 2}, we get the stronger statement
\begin{equation}
\label{eq: GB time delay is unresolvable2}
    \omega \abs*{\Delta T^{\text{GB}}} \sim \mu \omega b \left(\frac{r_g}{b}\right)^{D-1} \sim \frac{\omega}{b \Lambda^2} \left(\frac{r_g}{b}\right)^{D-3} \ll \left(\frac{r_g}{b}\right)^{D-3} \ll 1\,.
\end{equation}
It is now clear that the possible time advance induced by the GB term in the effective action is unresolvable, i.e. $\abs*{\Delta T^{\rm EFT}} \ll \omega^{-1}$, within the regime of validity of the EFT, regardless of how small the scale $\Lambda$, or the magnitude of the BH/spherical source.
In other words, Einstein-Gauss-Bonnet gravity does not violate infrared causality unless we choose to take it seriously in a regime in which it could not be viewed as having come from a consistent high-energy theory, irrespectively of what its precise completion is. This is consistent with the cosmological solutions considered in \cite{deRham:2020a}, and in a forthcoming work we will show the same result holds when considering multiple shock waves \cite{Shockwavepaper}.


\section{A casual case of cautionary acausality}
\label{sec: galileon toy model}

Our proposed ``infrared causality" condition would be useless were it not able to identify known field theories which violate causality. In this section we shall consider a scalar field theory in Minkowski space which is known to violate positivity bounds \cite{Tolley:2020gtv}, and thus is expected to violate standard causality conditions. In the next section we consider a related example accounting for gravity. Paralleling our discussion of the EFT of gravity, we will establish the regime of validity and compute the scattering time delay to demonstrate that this example does indeed lead to resolvable time advances within the regime of validity of the EFT.

\subsection{Quartic Galileon}
\label{subsec: galileon action}

We consider a scalar field invariant under $Z_{2}$: $\phi \rightarrow -\phi$ and Galilean shifts $\phi \rightarrow \phi + c + x^{\alpha}v_{\alpha}$ for some constant vector $v$.
The effective action up to $\mathcal{O}\big(\Lambda^{-8}\big)$ is
\begin{align}
\label{eq: galileon}
	S=\int & \d^{D}x\, \Lambda^{D-4}\bigg(-\frac{1}{2}\left(\partial\phi\right)^{2}+\frac{c_{1}}{\Lambda^{6}}\left(\partial\phi\right)^{2}\big([K^{2}]-[K]^{2}\big) \nonumber \\
	&+\frac{d_{1}}{\Lambda^{8}}[K^{4}]+\frac{d_{2}}{\Lambda^{8}}[K^{3}][K]+\frac{d_{3}}{\Lambda^{8}}[K^{2}][K]^{2}+\frac{d_{4}}{\Lambda^{8}}[K^{2}]^{2}+\frac{d_{5}}{\Lambda^{8}}[K]^{4} + \dots\bigg),
\end{align}
where we have denoted $K\mn=\partial_{\mu}\partial_{\nu}\phi$ and $[K]=K\indices{^{\alpha}_{\alpha}}$.
Note that we have not included the $\Lambda^{-2}$ contribution purely for convenience, and also chosen a non-canonical normalisation for the field. It is well-known that Galileons exhibit superluminalities \cite{Adams:2006sv,Goon:2010xh,deFromont:2013iwa}, although their relations with causality and consistent high-energy completions have a long history
\cite{Adams:2006sv,deRham:2013hsa,Keltner:2015xda,deRham:2017imi}.

To parallel our previous discussion of scattering time delays, we consider a spherically symmetric background $\bar{\phi}=\bar{\phi}(r)$. To order $\Lambda^{-8}$ the solution of the equations of motion is
\begin{align}
\label{eq: galileon back}
	\bar{\phi}(r)=&\frac{\alpha}{r^{D-3}}-c_{1}\frac{2(D-2)(D-3)^{4}}{(3D-5)r^{3D-5}}	\frac{\alpha^{3}}{\Lambda^{6}} \nonumber \\
	&+\frac{(D-3)^{3}(D-1)(D-2)\big[(20-8D)d_{1}-3(D-3)d_{2}-8(D-2)d_{4}\big]}{3r^{3(D-1)}}\frac{\alpha^{3}}{\Lambda^{8}},
\end{align}
where $\alpha$ is a constant measuring the scalar charge. The equation for perturbations is \eqref{eq: master eq tortoise} where now $W_{\ell}$ is given by equation \eqref{eq: Galileon Wl}.

\subsection{Regime of validity}
\label{subsec: galileon regime of validity bg}

We will now establish the EFT regime of validity for the background $\bar{\phi}$. Because of the Galileon symmetry, we need only consider operators built out of $K\mn = \partial_{\mu} \partial_{\nu} \phi$. The generic requirement for the traditional regime of validity of the effective theory is then
\be
\left( \frac{\partial}{\Lambda} \right)^p \ \left( \frac{K}{\Lambda^3}\right)^q \ll 1\,.
\ee
This closely parallels the effective theory of gravity where the Galileon invariant $K_{\mu\nu}$ plays the same role as the Riemann tensor. To leading order, $\bar{\phi} \sim \alpha/r^{D-3}$, and so demanding that this condition is valid at the point of closest approach $r \sim b$, we have
\begin{equation}
    \label{eq: galileon eft validity bg general}
    \frac{\alpha^q}{b^{p+q(D-1)}} \ll \Lambda^{p+3q}.
\end{equation}
The first, strongest bound come from $q \rightarrow \infty$
\begin{equation}
    \label{eq: galileon eft validity bg 1}
    b \gg r_{V}=\left(\frac{\alpha}{\Lambda^{3}}\right)^{\frac{1}{D-1}},
\end{equation}
where $r_{V}$ is known as the Vainshtein radius.
In addition, focusing on the bound $p \rightarrow \infty$ gives
\begin{equation}
    \label{eq: galileon eft validity bg 2}
    b \Lambda \gg 1 \, ,
\end{equation}
as in the case of gravity.

In order to establish the EFT regime of validity for the perturbations $\delta\phi$ around the background, we follow the same procedure and work in the limit of high momenta for fluctuations so that we may approximate $\partial_{\alpha}\delta\phi \sim i k_{\alpha}\delta\phi$. We then look for general scalar operators built out of the combinations
\be
\left( \frac{\partial}{\Lambda} \right)^p \ \left( \frac{K}{\Lambda^3}\right)^q \left( \frac{k}{\Lambda}\right)^r\ll 1\,.
\ee
As $k_{\alpha}k^{\alpha}=0$ to leading order, it is sufficient to focus on terms with the most $k$'s contracted with $K\mn$, specifically $k^{\mu} k^{\nu} K_{\mu \nu}$. For instance, we may focus on the set of operators of the form
\be
  \label{eq: galileon bound}
\Box^p [k^{\alpha} \partial_{\alpha}]^q \( k^{\mu} k^{\nu} K_{\mu \nu} \)^r \ll  \Lambda^{2p+2q+5r} \, .
\ee
Once again, since at leading order $\bar{\phi} \sim \alpha/r^{D-3}$, in the limit $q \rightarrow \infty$, \eqref{eq: galileon bound} gives
\begin{equation}
    \label{eq: galileon eft validity perts 2}
    \omega \ll \Lambda^{2} b.
\end{equation}
The limit $p \rightarrow \infty$ reproduces \eqref{eq: galileon eft validity bg 2}. The remaining non-trivial bound ($r \rightarrow \infty$) is
\be
|k^{\mu} k^{\nu} K_{\mu \nu}| \ll \Lambda^5 \, ,
\ee
which gives
\begin{equation}
    \label{eq: galileon eft validity perts 1}
    \omega^{2} \ll \frac{\Lambda^{5}b^{D-1}}{\alpha}  \, .
    \end{equation}
To summarise, equations \eqref{eq: galileon eft validity bg 1} and \eqref{eq: galileon eft validity bg 2} define the regime of validity for the backgrounds, and equations \eqref{eq: galileon eft validity perts 1} and \eqref{eq: galileon eft validity perts 2} define the regime of validity for the perturbations in our EFT.

\subsection{Scattering and time delay}
\label{subsec: galileon time delays}

Following the same procedure as in section \ref{subsec: phase shifts}, it is straightforward to compute the scattering time delay using \eqref{eq: Galileon Wl} which takes the form
\begin{align}
	\Delta T&=\bigg[-c_{1}\frac{1}{b^{-3+2D}}\frac{6\sqrt{\pi}(D-1)^{2}(D-3)^{2}\Gamma\big(D-\frac{3}{2}\big)}{\Gamma\big(D-1\big)}\frac{\alpha^{2}}{\Lambda^{6}} \nonumber \\
	&+(d_{1}+2d_{4})\frac{\omega^{2}}{b^{-3+2D}}\frac{3\sqrt{\pi}D(D-1)^{2}(D-3)^{2}\Gamma\big(D-\frac{3}{2}\big)}{\Gamma\big(D+1\big)}\frac{\alpha^{2}}{\Lambda^{8}}\bigg] \, ,
\end{align}
up to terms which are subleading in $1/\ell$ and $1/\Lambda$. We see that the first term leads to a time advance if $c_1>0$. Before concluding any violation of causality we must establish whether this is resolvable within the regime defined by the all four EFT bounds \eqref{eq: galileon eft validity bg 1}, \eqref{eq: galileon eft validity bg 2}, \eqref{eq: galileon eft validity perts 1} and \eqref{eq: galileon eft validity perts 2}. Indeed we have from \eqref{eq: galileon eft validity bg 1}
\be
\label{eq: deltaTc1}
|\omega \Delta T^{(c_1)}| \ll |c_{1} |\omega  b \, .
\ee
Imposing the condition \eqref{eq: galileon eft validity perts 2} then gives
\be
|\omega \Delta T^{(c_1)} |\ll |c_{1}| (b \Lambda)^2 \, .
\ee
Since none of the conditions impose an upper cutoff on $b$ (since a large $b$ pushes us towards the IR), we see that there is no difficulty in making the RHS as large as desired. Thus for $c_1>0$, this scalar model clearly violates causality.

Interestingly, on considering a combination of higher-dimension operators (those captured by the coefficient $d_{1}+2d_{4}$), we find that these make a comparable contribution to the time delay already when $\omega \sim \Lambda$, and so to infer any conclusion from the first term we need to impose
\begin{equation}
    \label{eq: galileon eft validity perts deltaT}
	\omega \ll \Lambda.
\end{equation}
Note that this does not imply a breakdown of the EFT at these scales, it rather reflects an accident of the leading Galileon terms being smaller than expected.  In other words, the EFT as a whole can still be under control when \eqref{eq: galileon eft validity perts deltaT} is violated, but the contribution from the quartic Galileon operator is simply suppressed as compared to that of other operators.
Regardless, even if we impose \eqref{eq: galileon eft validity perts deltaT} on \eqref{eq: deltaTc1}, we still have
\be
|\omega \Delta T^{(c_1)}| \ll |c_{1}| (b \Lambda) \, ,
\ee
for which the RHS remains arbitrarily large. This example illustrates that the resolvability criterion does not indiscriminately sweep all apparent acausalities under the carpet, but correctly identifies genuine ones.


\section{Gravitational positivity bounds from infrared causality}
\label{sec: positivity}

To clearly distinguish between the notions of asymptotic causality and infrared causality, we shall now argue that it is the latter condition which correctly reproduces the gravitational positivity bounds \cite{Caron-Huot:2021rmr} diagnosed in \cite{Alberte:2020jsk,Alberte:2020bdz,Alberte:2021dnj}. To this end, let us consider the Goldstone scalar model, now on a curved spacetime
\be\label{eq: Goldstone}
S = \int \d^D x \sqrt{-g} \left [- \frac{1}{2} (\nabla \phi)^2 + \frac{c}{\Lambda^{D}} ( \nabla \phi)^4 + \dots \right] \, .
\ee
To begin with, let us assume that the stress-energy of the scalar is sufficiently subdominant to other contributions that we can treat its backreaction on the metric perturbatively. We take the metric to be a general spherically symmetric form, (not necessarily $D$-dimensional Schwarzschild)
\be
\d s^2 = -B(r)^2 \d t^2+ A(r)^2 \d r^2+r^2 \d \Omega^2_{D-2} \, .
\ee
We can further assume that whatever sources the spherically symmetric solution, also sources a spherically symmetric background configuration for the scalar. At leading order, this background is determined by
\be\label{eq: background soln}
\bar{\phi}'(r) = \frac{\alpha}{r^{D-2} C(r) } + {\cal O}(\Lambda^{-D}) \, ,
\ee
with $C(r)=B(r)/A(r)$.

Considering fluctuations $\phi= \bar \phi + \delta \phi$, defining $\chi = r^{(D-2)/2} \sqrt{C} \( 1- \frac{6 c \alpha^2}{ \Lambda^{D} r^{2D-4} A^2 C^2} \) \delta \phi$, and performing the Langer transformation $r= e^{\rho}$, we have for each partial wave
\be
\chi_{\ell}''(\rho) + W_{\ell}(\rho) \chi_{\ell}(\rho) =0\,,
\ee
where
\be
W_{\ell}(\rho) =\frac{e^{2 \rho}}{C^2} \omega^2 \( 1+ \frac{8 c \alpha^2}{\Lambda^{D} r^{2D-4} A^2 C^2} \)-A^2 \omega^2 b^2 \( 1+ \frac{8 c \alpha^2}{\Lambda^{D} r^{2D-4} A^2 C^2} \)+\dots \, ,
\ee
up to subleading terms negligible at large $\ell$, and terms of higher order in $1/\Lambda$.

Given the turning point defined by $W_{\ell}(\rho_t)=0$, it is helpful to split $W_{\ell}$ as $W_{\ell}=\omega^2 \(U_{\ell}^{(1)}+\frac{c \alpha^2}{\Lambda^{D-2} } U_{\ell}^{(2)}\)$ where
\be
U_{\ell}^{(1)} (\rho,\rho_t)= \( \frac{e^{2 \rho}}{C^2(\rho)} -\frac{e^{2 \rho_t}}{C^2(\rho_t)} \)  - b^2\(A^2(\rho)-A^2(\rho_t)\)  \, ,
\ee
and
\be
U_{\ell}^{(2)} (\rho,\rho_t)= 4 \(  \frac{e^{-(2D-3) \rho}}{A^2 (\rho) C^3(\rho)} - \frac{e^{-(2D-3) \rho_t}}{ A^2(\rho_t) C^3(\rho_t)}   \) -  4 b^2 \( \frac{e^{-(2D-4) \rho} }{ C^2(\rho)} -\frac{e^{-(2D-4) \rho_t} }{ C^2(\rho_t)}  \)    \, .
\ee
The net time delay at fixed impact parameter is given up to order $1/\Lambda^{D}$ by
\be
\Delta T^{\rm net}_b = 2 \int_{\rho_t}^{\infty} \(\sqrt{U_{\ell}^{(1)} (\rho,\rho_t)} -e^\rho\) \d \rho -r_t+ \frac{c \alpha^2}{ \Lambda^{D-2} }   \int_{\rho_t}^{\infty} \frac{U_{\ell}^{(2)} (\rho,\rho_t)}{\sqrt{U_{\ell}^{(1)} (\rho,\rho_t)}} \d \rho+ \dots \, .
\ee
We now define the $\text{GR}$ time delay as the time delay we would obtain in the limit $\Lambda \rightarrow \infty$, in which all corrections from heavy modes are removed. Since the turning point $\rho_t$ in general depends on $\Lambda$, this will be written in terms of the turning point $\rho_t^0$, for which $B(\rho_t^0) b=e^{ \rho_t^0}$.
Thus
\be \label{eq: GRdelay}
\Delta T^{\rm GR}_b = \lim_{\Lambda \rightarrow \infty} \Delta T^{\rm net}_b = 2 \int_{\rho_t^0}^{\infty} \(\sqrt{U_{\ell}^{(1)} (\rho,\rho_t^0)} -e^{\rho}\) \d \rho -r_t^0 \, .
\ee
Putting this together, the EFT time delay at fixed impact parameter is
\ba
\Delta T^{\rm EFT}_b &=&  \frac{c \alpha^2}{\Lambda^{D} }   \int_{\rho^0_t}^{\infty}  \frac{U_{\ell}^{(2)} (\rho,\rho^0_t)}{\sqrt{U_{\ell}^{(1)} (\rho,\rho^0_t)}} \d \rho
 + 2 \int_{\rho_t}^{\infty} \sqrt{U_{\ell}^{(1)} (\rho,\rho_t)}  \d \rho - 2 \int_{\rho_t^0}^{\infty} \sqrt{U_{\ell}^{(1)} (\rho,\rho_t^0)}  \d \rho +\cdot \nn \\
 &=& \frac{c \alpha^2}{\Lambda^{D} }   \int_{\rho^0_t}^{\infty}  \frac{U_{\ell}^{(2)} (\rho,\rho^0_t)}{\sqrt{U_{\ell}^{(1)} (\rho,\rho^0_t)}} \d \rho
 + (\rho_t-\rho_t^0) \int_{\rho^0_t}^{\infty} \frac{\p_{\rho_t^0}U_{\ell}^{(1)} (\rho,\rho^0_t)}{\sqrt{U_{\ell}^{(1)} (\rho,\rho^0_t)}}  \d \rho  +\cdots \,.
\ea
So far, preserving causality seems to imply $\Delta T^{\rm EFT}_b>0$, which would demand $c>0$, consistent with known positivity bounds \cite{Pham:1985cr,Ananthanarayan:1994hf,Adams:2006sv}.

\subsection{Bound from asymptotic causality}

As already illustrated in section~\ref{sec: galileon toy model},
causality can be used to put bounds on the Wilson coefficients in an effective theory. At face value, in the case of the prototypical Goldstone EFT  \eqref{eq: Goldstone}, causality simply seems to indicate $c>0$, however we shall see that this bound can actually be slightly violated in the presence of gravity.

Crucially, we now include the gravitational backreaction of the field $\bar \phi$. We remain at impact parameters for which the weak gravitational field approximation is valid, so when including the scalar field backreaction we can take $A\approx B \approx C\approx 1$. The metric in harmonic gauge then satisfies
\be
\Box \(h_{\mu \nu} - \frac{1}{2} \eta_{\mu\nu} h \) = -\frac{2}{\mpl^{D-2}} T_{\mu\nu} \, .
\ee
This is easily solved for a spherically symmetric source \eqref{eq: background soln}, and schematically we have\footnote{For the remainder of this section we omit order unity factors as our goal is to estimate orders of magnitudes. With this in mind, the arguments in this and the following section apply equally well to the time delay at fixed $b$ or $\ell$.}
\be\label{eq: weak gravity}
h \sim \frac{\alpha^2}{\mpl^{D-2} r^{2(D-3)}} \, .
\ee
It is straightforward to estimate the GR time delay \eqref{eq: GRdelay} from the backreaction of the scalar as
\be
\Delta T^{\phi, \text{GR}} \sim \frac{\alpha^2}{\mpl^{D-2} b^{2(D-3)}}  b\, .
\ee
This adds to the contribution from any other gravitational source such as the Schwarzschild background due to a localised source of mass $M$ with Schwarzschild radius $r_g^{D-3} \sim M/\mpl^{D-2}$
\be
\Delta T^{M, \text{GR}}\sim \frac{r_g^{D-3}}{b^{D-4}} \, ,
\ee
while in the weak field limit, the EFT contribution is essentially dominated by its value in Minkowski spacetime
\be
\Delta T^{\rm EFT} \sim \frac{c \alpha^2}{ \Lambda^{D} b^{2D-5}} \, .
\ee
Putting this together, the condition for asymptotic causality, i.e. the absence of any resolvable time advance is
\be
\Delta T^{\text{net}} \sim \frac{r_g^{D-3}}{b^{D-4}} +\frac{\alpha^2}{\mpl^{D-2} b^{2D-6}}  b \( 1 + \frac{c \mpl^{D-2} }{\Lambda^D b^2} \) \gtrsim -\omega^{-1}\,.
\ee
If $c>0$, the net time delay is always positive, but for $c<0$ the net delay can in principle always be made negative by choosing $b$ small enough.
More precisely we can read this equation as a lower bound on the coefficient $c$, similar in spirit to the positivity bounds
\be\label{eq: cbound1}
c \gtrsim -\frac{\Lambda^D b^2}{\mpl^{D-2}} -  \frac{\Lambda^D b^{D-1} r_g^{D-3}}{\alpha^2} -\frac{\Lambda^D b^{2D-5}}{\alpha^2 \omega} \, .
\ee
Our goal is to find the tightest version of this bound for which the RHS is as large as possible. We may always choose $r_g=0$, or $\alpha$ large enough so that the Schwarzschild contribution is negligible. The largest $\omega$ we are allowed is $\Lambda^2 b$, and the largest $\alpha$ can be fixed from requiring $(\partial \phi)^2 \lesssim \Lambda^D$, so that $\alpha \lesssim \Lambda^{D/2} b^{D-2}$. With these choices we have
\be
c \gtrsim -\frac{\Lambda^D b^2}{\mpl^{D-2}} -\frac{1}{\Lambda^2 b^2} \, .
\ee
Extremising the RHS gives $b=\Lambda^{-1} \( \frac{\mpl}{\Lambda}\)^{(D-2)/4} \gg \Lambda^{-1}$ which in turn gives the bound
\be \label{eq: asympbound}
c \gtrsim -  \( \frac{\Lambda}{\mpl}\)^{(D-2)/2}  \, .
\ee
This is the most optimistic bound\footnote{If we had fixed the maximum $\alpha$ by requiring the weak gravitational field approximation is still valid, i.e. $\alpha^2 \sim \mpl^{D-2} b^{2D-6}$ then substituting in \eqref{eq: cbound1} with $r_g=0$ and the maximum $\omega$ gives
$c \gtrsim -\Lambda^D b^2\mpl^{2-D} -\Lambda^{D-2} \mpl^{2-D}$.
Then demanding that $(\partial \phi)^2 \ll \Lambda^D$ gives $b^2 \gg \Lambda^{-1} \(\mpl/\Lambda \)^{D-2} $, leading to the even weaker statement $c \gtrsim -1$.} we can get from this analysis, and occurs at $h \sim (\Lambda/\mpl)^{(D-2)/2} \ll 1$, well within the weak gravitational field approximation.

Crucially, the bound \eqref{eq: asympbound} is weaker than the gravitational positivity bounds \cite{Caron-Huot:2021rmr} observed in \cite{Alberte:2020jsk,Alberte:2020bdz,Alberte:2021dnj}. Indeed we would never expect positivity bounds to give a condition of the form \eqref{eq: asympbound} due to the non-analytic square root of the gravitational coupling constant.

\subsection{Positivity bound from IR causality}

If we now consider the infrared causality condition, a much clearer picture emerges. Again working in the weak gravitational field regime, we focus now only on the EFT time delay
 \be
\Delta T^{\rm EFT} \sim \frac{c \alpha^2}{\Lambda^{D} b^{2D-5}} \, .
\ee
In order to ask whether this is resolvable within in the regime of the EFT we again use the maximum allowed frequency $\omega \ll \Lambda^2 b$ so that
\be
\label{condition1}
\omega \Delta T^{\rm EFT} \ll \frac{c \alpha^2}{\Lambda^{D-2} b^{2D-6}} \, .
\ee
Now, in the absence of gravity, the only clear upper bound we can impose on $\alpha$ comes from demanding $(\partial \phi)^2 \ll \Lambda^{D}$ which would tell us
\be
\omega \Delta T^{\rm EFT} \ll c (b \Lambda)^2 \, .
\ee
Since the RHS can be arbitrarily large we find that if $c<0$ we can establish an arbitrarily large resolvable time advance, regardless of how small the magnitude of $c$ is. This is of course consistent with positivity bounds in Minkowski spacetime that demand the strict requirement \cite{Pham:1985cr,Ananthanarayan:1994hf,Adams:2006sv}
\be
c >0 \, .
\ee
The central difference when we include gravity is that, as we increase the scale $\alpha$ in the field profile, there comes a point at which we can no longer neglect the gravitational backreaction of the scalar field $\phi$ itself. Demanding that the backreaction is under control, which was implicit in our calculation so far, amounts to
\be
\frac{\alpha^2}{\mpl^{D-2} b^{2D-6}} \ll 1 \, .
\ee
Putting this condition into the \eqref{condition1} we infer that
\be
\omega \Delta T^{\rm EFT}_b  \ll \frac{c\mpl^{D-2}  }{\Lambda^{D-2}}\,.
\ee
Now the condition for infrared causality, $ \Delta T^{\rm EFT}_b \gtrsim -\omega^{-1}$, becomes the statement
\be
c \gtrsim - \frac{\Lambda^{D-2}}{\mpl^{D-2}} \, ,
\ee
which is (up to order unity factors) exactly the gravitational positivity bound derived in \cite{Caron-Huot:2021rmr} and conjectured in \cite{Alberte:2020jsk} (for the case of $D=4$).

We learn something very important from the above analysis. The fact that the coefficient $c$ is allowed to be slightly negative when we include gravity has nothing whatsoever to do with the positive contribution from $\Delta T^{\rm GR}$. It is entirely down to the negative sign in the resolvability criterion $\Delta T^{\rm EFT} \gtrsim -\omega^{-1}$. This is the most compelling evidence that causality in a gravitational theory is determined by the infrared causality condition and not the asymptotic causality condition. A similar observation was made in the EFT of gravity in $D=4$ in \cite{Junpaper}. There, it was noted that if we only impose the criterion of asymptotic causality, we allow for values for the Wilson coefficients which are in conflict with positivity bounds. In a forthcoming work \cite{Shockwavepaper} we will show that all these conclusions are paralleled for the case of shock waves, including the multiple shock wave solutions considered in \cite{Camanho:2014apa} (see also \cite{Edelstein:2016nml,Kologlu:2019bco,AccettulliHuber:2020oou,Ge:2020tid,Edelstein:2021jyu}). A different way of interpreting these results is that if we take for granted the gravitational positivity bounds from the outset, then it is clear that we can never generate a resolvable time advance within the regime of validity of the EFT.


\section{Discussion}
\label{sec: discussion}
In this work we have highlighted how to understand causality within a given low-energy effective theory without appealing to its precise UV completion. We have distinguished between two notions of causality, both of which can be defined from the S-matrix via the scattering phase shift. Asymptotic causality demands that there is no resolvable net time advance, and in a Lorentz invariant theory may be regarded as the causality set by the asymptotic Minkowski geometry. Infrared causality demands that there is no resolvable time advance relative to the GR background which is common to all interacting states due to the equivalence principle. The latter condition is a stronger one, since the former one allows superluminal propagation with respect to the metric which sets null geodesics. We find that in known examples of truncated EFTs which admit a standard UV completion, such as the Einstein-Gauss Bonnet theory \eqref{eq: effective action} or the Goldstone model \eqref{eq: Goldstone}, infrared causality is automatically respected in the regime of validity of the effective theory. In this sense causality of consistent low-energy effective theories, can be understood without appealing to the precise UV completion, be it an infinite tower of higher-spins or loop effects.

In the case of low-energy EFTs that consistently derive from integrating out heavy modes of a standard (partial) UV completion, the apparent superluminal propagation indicated by a purely classical analysis can be shown to never be resolvable, due to the cutoff of the energy of scattered states required for validity of the EFT. To illustrate the emergence of a high-energy cutoff, we explicitly compute the effect of higher-dimension operators in the EFT expansion. Specifically, in the case of Einstein-Gauss-Bonnet gravity, corrections from dimension-8 operators already induce higher-order frequency dependence in the propagation equations which impose a cutoff which is sufficient to ensure infrared causality. By contrast, known EFTs that violate positivity bounds can be shown to lead to resolvable time advances, relative to the GR background.

We show that demanding infrared causality for spherically symmetric scattering on a simple scalar Goldstone model essentially imposes the known gravitational positivity bounds. Equivalently stated, if we impose positivity bounds on the EFT from the outset, it is impossible to generate a resolvable time advance for scattering (relative to the background metric) regardless of the mass of the source. This is consistent with previous analyses that included cosmological backgrounds \cite{deRham:2020a}, and in a forthcoming work we will show the same is true for scattering across multiple shock waves \cite{Shockwavepaper}. By contrast, the weaker condition of asymptotic causality would allow for Wilson coefficients which violate known positivity bounds, as noted also in \cite{Junpaper}.
Our results support the notion that infrared causality is the most pragmatic way to understand how causality is realised entirely within the low-energy EFT. Furthermore, it gives a condition which is meaningful for both tree-level weakly coupled completions where the higher-dimension operators primarily come from integrating out higher-spin states, or more general cases where the higher-dimension operators are dominated by loop contributions, and when the UV completion is itself strongly coupled.


\acknowledgments
The authors  would like to acknowledge the use of the \textit{xTras} package of \textit{xAct} for Mathematica \cite{martin2002xact, Nutma:2013zea}.
The work of AJT and CdR is supported by STFC grants ST/P000762/1 and ST/T000791/1. CdR thanks the Royal Society for support at ICL through a Wolfson Research Merit Award.  CdR is supported by the European Union Horizon 2020 Research Council grant 724659 MassiveCosmo ERC2016COG. CdR is also supported by a Simons Foundation award ID 555326 under the Simons Foundation Origins of the Universe initiative, Cosmology Beyond Einstein's Theory and by a Simons Investigator award 690508. AJT thanks the Royal Society for support at ICL through a Wolfson Research Merit Award. CC and AM are funded by the President's PhD Scholarships.


\appendix

\section{Master variables for metric perturbations}
\label{app: master variables}
In this appendix, we summarise the procedure laid out in \cite{Kodama:2003a} for identifying the propagating degrees of freedom (the ``master variables") of metric perturbations, and apply it to our case of a $D$-dimensional Schwarzschild BH in the EFT of gravity.

Following the notation of \cite{Kodama:2003a}, we write the background spacetime as
\begin{equation}
    \d s^2 = \eta_{ab}(y) \d y^a \d y^b + r^2 \gamma_{ij} \d z^i \d z^j,
\end{equation}
where $\eta_{ab}$ is the Lorentzian metric of the 2-dimensional orbit spacetime:
\begin{equation}
    \eta_{ab}(y) \d y^a \d y^b = -f(r) \d t^2 + \frac{1}{f(r)} \d r^2,
\end{equation}
and $\gamma_{ij}$ is the metric on the unit $(D-2)$-sphere $S^{D-2}$.
The metric perturbations follow the same index conventions as the background metric.
Furthermore, we denote the covariant derivative with respect to $\gamma_{ij}$ on $S^{D-2}$ by $\hat{D}_j$, and the associated Laplace-Beltrami operator by $\LB \equiv \gamma^{ij}\hat{D}_i\hat{D}_j$.
The covariant derivative on the orbit spacetime is denoted by $D_a$, and the associated Laplace-Beltrami operator by $\Box_2 \equiv \eta^{ab}D_aD_b$.
The Laplace-Beltrami operator on the full spacetime manifold is denoted simply by $\Box \equiv g^{\alpha \beta}\nabla_{\alpha}\nabla_{\beta}$.
\subsection{Tensor modes}
\label{subapp: tensor modes}
The tensor-type metric perturbations are expanded in terms of tensor spherical harmonics $\TT_{ij}$ on $S^{D-2}$, which satisfy
\begin{subequations}
\label{eq: tensor spherical harmonics}
\begin{align}
    &\left(\LB + k_T^2\right)\TT_{ij} = 0,\\
    &\TT\indices{^i_i} = 0, \quad \hat{D}_j\TT\indices{^j_i} = 0,
\end{align}
\end{subequations}
where $k_T^2$ is the eigenvalue of $\LB$ acting on the tensor $\TT_{ij}$, and takes discrete values:
\begin{equation}
    k_T^2 = \ell(\ell+D-3)-2, \quad \ell = 1, 2, \dots.
\end{equation}
All mode numbers (e.g. $\ell$) that could label $\TT_{ij}$ have been suppressed.
Expressions for the symmetric tensor spherical harmonics on the $(D-2)$-sphere in terms of the mode number $\ell$ can be found in \cite{Higuchi:1987} but are not relevant for our purposes.
For each such tensor, the tensor-type metric perturbations can be written at each $\ell$ as
\begin{equation}
\label{eq: tensor type perturbations}
    h_{ab} = 0, \quad h_{ai} = 0, \quad h_{ij} = 2 r^2 H_T \TT_{ij}.
\end{equation}
It is clear that the only tensor-mode freedom is in the function $H_T \equiv H_T(t,r)$.
The dynamics of this scalar function are determined by the first-order perturbation of the Einstein-Gauss-Bonnet equations \eqref{eq: EGB eqs}.
To obtain the master equation in the form \eqref{eq: master eq}, it is simply a matter of rescaling $H_T$ accordingly:
\begin{equation}
    \Phi_T = r^{(D-2)/2}\left(1 - 4 (D-4) \cGB \mu \rgr^{D-1} \right)H_T.
\end{equation}

\subsection{Vector modes}
\label{subapp: vector modes}
The vector-type metric perturbations are expanded in terms of vector spherical harmonics $\VV_i$ on $S^{D-2}$, which satisfy
\begin{subequations}
\label{eq: vector spherical harmonics}
\begin{align}
    &\left(\LB + k_V^2\right)\VV_{i} = 0,\\
    &\hat{D}_j\VV^j = 0,
\end{align}
\end{subequations}
where $k_V^2$ is the eigenvalue of $\LB$ acting on the vector $\VV_i$ and takes discrete values\footnote{The $\ell=1$ vector harmonic corresponds to rotational perturbations of the BH and not a dynamical degree of freedom.}:
\begin{equation}
    k_V^2 = \ell(\ell+D-3)-1, \quad \ell = 1, 2, \dots.
\end{equation}
All mode numbers (e.g. $\ell$) that could label $\VV_i$ have been suppressed.
For each vector, the vector-type metric perturbations can be written at each $\ell$ as
\begin{equation}
\label{eq: vector type perturbations}
    h_{ab} = 0, \quad h_{ai} = r f_a \VV_i, \quad h_{ij} = 2 r^2 H_T \VV_{ij},
\end{equation}
where
\begin{equation}
    \VV_{ij} = -\frac{1}{2k_V}\left(\hat{D}_i\VV_j + \hat{D}_j\VV_i\right).
\end{equation}
Identifying the master variable for vectors is a more involved process than it was for tensors because of the extra variables $f_a$ in the initial parameterisation of the metric perturbations \eqref{eq: vector type perturbations}.
Following \cite{Kodama:2003a} we begin by constructing the gauge-invariant variables
\begin{equation}
    F_a = f_a + \frac{r}{k_{V}} D_a H_T.
\end{equation}
The master variable $\Phi_T$ is directly related to $F_r$ as
\begin{equation}
    \Phi_V = r^{-(D-6)/2}\left(1 - 4(D-4) \cGB \mu \rgr^{D-1} \right)   f(r) F_r.
\end{equation}
\subsection{Scalar modes}
\label{subapp: scalar modes}
The scalar-type metric perturbations are expanded in terms of scalar spherical harmonics $\Scalar$ on $S^{D-2}$, which satisfy
\begin{equation}
    \left(\LB + k_S^2\right)\Scalar = 0,
\end{equation}
where $k_S^2$ is the eigenvalue of $\LB$ acting on the scalar $\Scalar$ and takes discrete values\footnote{The $\ell=0$ scalar harmonic corresponds to a shift in the BH mass, while the $\ell=1$ scalar harmonic turns out to be pure gauge \cite{Kodama:2003a}, so neither are dynamical.}:
\begin{equation}
    k_S^2 = \ell(\ell+D-3), \quad l = 0,1,2,\dots\, .
\end{equation}
All mode numbers (e.g. $\ell$) that could label $\Scalar$ have been suppressed, as before.
For each $\ell$, the scalar-type metric perturbations can be written as
\begin{equation}
\label{eq: scalar type perturbations}
    h_{ab} = f_{ab} \Scalar, \quad h_{ai} = r f_a \Scalar_i, \quad h_{ij} = 2 r^2 \left(H_{\ell} \gamma_{ij} \Scalar + H_T \Scalar_{ij}\right),
\end{equation}
where
\begin{subequations}
\begin{align}
    &\Scalar_i = -\frac{1}{k_S}\hat{D}_i\Scalar,\\
    &\Scalar_{ij} = \frac{1}{k_S^2}\hat{D}_i\hat{D}_j\Scalar + \frac{1}{D-2}\gamma_{ij}\Scalar.
\end{align}
\end{subequations}
As for the vectors, identifying the master variable in terms of those in \eqref{eq: scalar type perturbations} becomes easier after constructing the following gauge-invariant variables:
\begin{subequations}
\begin{align}
    &F = H_{\ell} + \frac{1}{D-2}H_T + \frac{1}{r}D^a r X_a,\\
    &F_{ab} = f_{ab} + D_a X_b + D_b X_a,\\
    &X_a = \frac{r}{k_S}\left(f_a + \frac{r}{k_S}D_a H_T\right).
\end{align}
\end{subequations}
There is a change of variables of the form
\begin{equation}
    F = \alpha(r) \Phi_S + \beta(r) \partial_r \Phi_S,
\end{equation}
such that all components of the perturbed Einstein-Gauss-Bonnet equations are automatically satisfied when $\Phi_S$ obeys the master equation \eqref{eq: master eq}.
To leading order in $\mu$, we need
\ba
\alpha(r)&=&\frac{1}{4(D-2)r^{(D-2)/2}H}\bigg\{4k_{S}^{4}+2(D-2)(D-6)k_{S}^{2}-2(D-2)^{2}(D-4) \\	&+&\big[6(D-2)k_{S}^{2}-(D-2)^{2}\left(D(D-5)+10\right)\big]\rgr^{D-3}+(D-1)(D-3)^{3}\rgr^{2(D-3)}\bigg\} \nonumber \\
&+&\cGB\mu \frac{D-4}{4(D-2)r^{\frac{D}{2}-1}H^{2}}\bigg\{16k_{S}^{6}-8(D-2)\left(D(2D-7)+12\right)k_{S}^{4}  \nonumber \\ &+&16(D-2)^{4}\left(D(2D-7)+9\right)k_{S}^{2}-8(D-2)^{3}\left(D(2D-7)+8\right)\nonumber \\
&+&\big[4(D-2)\left(D(5D-17)+24\right)k_{S}^{4} -24(D-2)^{2}\left(D(3D-10)+11\right)k_{S}^{2} \nonumber \\
	&&\hphantom{\big[4(D-2)\left(D(5D-17)+24\right)k_{S}^{4}} +4(D-2)^{3}\left(D(13D-43)+42\right)\big]\rgr^{D-3} \nonumber \\ &+&\big[16(D-1)(D-2)^{2}(2D-3)k_{S}^{2}-2(D-1)(D-2)^{3}\left(D(3D+7)-18\right)\big]\rgr^{2(D-3)} \nonumber \\ &-&\big[(D-1)^{2}(D-2)^{3}\left(D(D-11)+14\right)\big]\rgr^{3(D-3)}\bigg\}\,,\nn
\ea
where
\begin{equation}
    H= k_{S}^{2} - D + 2 + \frac{(D-1)(D-2)}{2} \rgr^{D-3},
\end{equation}
and
\begin{equation}
	\beta(r)=\frac{1}{r^{(D-4)/2}}\bigg[1-\rgr^{D-3}+4(D-4)\cGB\mu \rgr^{D-1}\bigg(1+\frac{D-5}{2}\rgr^{D-3}\bigg)\bigg].
\end{equation}
\section{Potentials}
\label{app: potentials}
Here we collect the potentials for the tensor, vector and scalar modes to first order in the EFT expansion parameter $\mu$.
They appear in the master equations \eqref{eq: master eq} for their respective modes.
For the tensor potential, we find
\ba
\label{eq: tensor potential app}
    \frac{V_{T}}{f} &=& \frac{1}{r^2}\bigg[k_{T}^2\left(1 + 8 \cGB \mu (D-1) \rgr^{D-1}\right) \\
    &+& \frac{D(D-6)+16}{4}\left(1 - 32\cGB \mu \frac{(D-1)(D-6)}{D(D-6)+16} \rgr^{D-1}\right) \nn\\
    &+& \frac{(D-2)^{2}}{4}\rgr^{D-3}\left(1 - 2\cGB \mu \frac{(D-4)\left[3D(D-3)(D-6)-32\right]}{(D-2)^{2}}\rgr^{D-1}\right)\bigg]. \nonumber
\ea
We can identify the tensor angular speed from the coefficient of the $k_T^2/r^2$-term in \eqref{eq: tensor potential app}:
\begin{equation}
    v_{\Omega,T}^2 = 1 + 8 \cGB \mu (D-1) \rgr^{D-1}.
\end{equation}
For the vector potential, we get
\ba
\label{eq: vector potential app}
\frac{V_{V}}{f} &=&
\frac{1}{r^2}\bigg[k_{V}^2\left(1 - 4(D-1)(D-4) \cGB \mu \rgr^{D-1}\right)  \\
    &+& \frac{D(D-6)+12}{4}\left(1 + 16 \frac{(D-1)(D-4)(3D-5)}{D(D-6)+12} \cGB\mu \rgr^{D-1}\right)\nn \\
    &-& \frac{3(D-2)^{2}}{4}\rgr^{D-3}\left(1 - \frac{2(D-4)\big[D^2(5D-57)+134D-88\big]}{3(D-2)^{2}}\cGB \mu \rgr^{D-1}\right)\bigg]. \nonumber
\ea
We can identify the vector angular speed from the coefficient of the $k_V^2/r^2$-term in \eqref{eq: vector potential app}:
\begin{equation}
    v_{\Omega,V}^2 = 1 - 4(D-1)(D-4) \cGB \mu \rgr^{D-1}.
\end{equation}
Finally, for the scalar potential we have
\ba
\label{eq: scalar potential app}
\frac{V_{S}}{f}=	\frac{1}{16H^{2}r^{2}} \mathcal{V}^{\rm GR}
+\cGB \mu \frac{D-4}{16H^{3}r^{2}}\rgr^{D-1}\mathcal{V}^{\rm GB}\,,
\ea
where the contribution from GR is given by
\ba
\mathcal{V}^{\rm GR}
&=&4\left(k_{S}^{2}-D+2\right)^{2}\left(4k_{S}^{2}+D^{2}-6D+8\right) \\	&-&12(D-2)\left(k_{S}^{2}-D+2\right)\big[(D-6)k_{S}^{2}+D^2(D-8)+22D-20\big]\rgr^{D-3} \nonumber \\ &+&
4k_{S}^{2}(D-1)(D-2)(18+D(2D-11))\rgr^{2(D-3)} \nn\\
&+&
(D-1)(D-2)^2(D(D-6)(D-13)-96)
\rgr^{2(D-3)} \nonumber \\
&+&(D-1)^{2}(D-2)^{4}\rgr^{3(D-3)}\,,\nn
\ea
and that from the GB term by
\ba
\mathcal{V}^{\rm GB}&=&
-128 (D-1)(k_{S}^2-D+2)^3\big[(D-2)^{2} + k_{S}^{2}\big]  \\
&-&8k_{S}^{2}(D-2)(k_{S}^{2}-D+2)^{2} (D(21D-131)+140) \rgr^{D-3} \nonumber \\
&+&8(D-2)^2(k_{S}^{2}-D+2)^{2} (8D^3-125 D^2+347 D-260)\rgr^{D-3} \nonumber \\
&-&36k_{S}^{2}(D-1)(D-2)^{2}(k_{S}^{2}-D+2)(D-3)(D-12)  \rgr^{2(D-3)} \nonumber \\	
&+&12(D-1)(D-2)^{3}(k_{S}^{2}-D+2) \(3D^{2}-77 D+140\)\rgr^{2(D-3)}\nonumber \\ &+&2k_{S}^{2}(D-1)^{2}(D-2)^{2}\left(112-9D^3+97D^2-226D\right) \rgr^{3(D-3)} \nonumber \\ &+&2(D-1)^{2}(D-2)^{3}\(9D^{3}-113 D^2+258 D-128\) \rgr^{3(D-3)} \nonumber \\ &-&(D-1)^{3}(D-2)^{3}\big[3D(D-3)(D-6)-32\big]\rgr^{4(D-3)}\,.\nn
\ea
Due to the complicated dependence of the scalar potential on the eigenvalue $k_S^2$, the scalar angular speed cannot be readily extracted from \eqref{eq: scalar potential app} unless one takes the large $\ell$ ($k_S$) limit.

\section{Easy as (dimension-) 4, 6, 8}
\label{app: higher dim ops}
We will now outline why we considered the operator in \eqref{eq: S8 extension} as the next term in our EFT expansion. The EFT contains operators with arbitrary number of derivatives. In what follows, we shall focus on the higher-derivative terms in the equations of motion, as those are the ones that will lead to the most stringent bound on $\omega$, and may therefore neglect non-derivative terms. In practice, this means we can freely replace covariant with partial derivatives ($\nabla \rightarrow \partial$). Moreover, since we are implementing a perturbative approach, we may freely use the GR equations of motion for the tensor fluctuations, namely $\Box h \approx 0$, in the EFT corrections.

For this section, we are specifically interested in the dynamics of tensor modes.
Recall that we denote base space indices by $a,b,c,d,\dots \in \{0,1\}$, and orbit space indices by $i,j,k,l,\dots \in \{2,\dots,D-1\}$.
For the tensor perturbations, $h_{ij}$ are the only non-vanishing components.
They therefore inherit transverse-tracelessness from the tensor spherical harmonics, i.e. $\nabla_{\alpha}h\indices{^\alpha_\mu} = 0$ and $h\indices{^\alpha_\alpha} = 0$.

\subsection{Dimension-4 operators}
\label{subapp: dim 4 ops}
First, consider the dimension-4 operators. Schematically, i.e. suppressing the indices,
\begin{equation}
    \mathcal{L}_{\text{D4}} \sim \frac{R^{2}}{\Lambda^2}.
\end{equation}
As previously discussed in section \ref{sec: BH in EFT}, $R^{2}/\Lambda^2$ and $R_{\mu\nu}^{2}/\Lambda^2$ contribute to the perturbed Einstein equations at $\mathcal{O}(\mu^{3})$, which are higher-order than the leading EFT contributions.

Even though we can compute the equations of motion, it will be instructive to just construct the possible terms in them by other means.
In particular, we know from dimensional analysis that $\mathcal{L}_{\text{D4}}$ will introduce terms of the schematic form
\begin{equation}
    \nabla^{4}h, \quad R\nabla^{2}h, \quad R^{2}h.
\end{equation}
Of these, only the first term can possibly provide us with four derivatives.
Leaving two indices free, the only possible contractions (up to commutation of covariant derivatives, which we ignore) are
\begin{equation}
    \nabla^{4}h \quad \rightarrow \quad \Box^{2}h_{\mu\nu}, \quad \Box\nabla_{\mu}\nabla_{\alpha}h\indices{^{\alpha}_{\nu}}, \quad \Box\nabla_{\mu}\nabla_{\nu}h\indices{^{\alpha}_{\alpha}}.
\end{equation}
Notably, we are always forced to contract at least two covariant derivatives with each other.
We can see that none of the above terms are truly higher derivative in nature upon substituting $\Box \approx 0$ up to non-derivative terms at leading order.

\subsection{Dimension-6 operators}
\label{subapp: dim 6 ops}
Knowing that dimension-4 operators do not result in higher-derivative equations, we now turn our attention to dimension-6 operators.
Dimensional analysis means that $\mathcal{L}_{\text{D6}}$ will introduce terms in the perturbation equations of motion of the schematic form
\begin{equation}
    \nabla^{6}h, \quad R\nabla^{4}h, \quad R^{2}\nabla^{2}h, \quad R^{3}h.
\end{equation}
The first term inevitably has two covariant derivatives contracted with each other, giving a $\Box \approx 0$, so we only need to consider the second term.
Next, since the tensor modes are transverse and traceless, the only seemingly non-trivial contraction of indices (up to symmetries of the Riemann tensor) is
\begin{equation}
    \label{eq: Dim 6}
    R\nabla^{4}h \rightarrow R^{\alpha\rho\beta\sigma}\nabla_{\mu}\nabla_{\nu}\nabla_{\alpha}\nabla_{\beta}h_{\rho\sigma}.
\end{equation}
To find the highest-derivative term, we trade in covariant with partial derivatives and using the form of the Riemann tensor on the Schwarzschild background
\begin{subequations}
\label{eq: Riemann components}
\begin{align}
    &R^{abcd} = -\frac{f''}{2}\left(g^{ac}g^{bd} - g^{ad}g^{bc}\right), \\
    &R^{aibj} = -\frac{f'}{2 r}g^{ab}g^{ij}, \\
    &R^{ijkl} = \frac{1-f}{r^{2}}\left(g^{ik}g^{jl} - g^{il}g^{jk}\right),
\end{align}
\end{subequations}
we find that
\begin{align}
	R\indices{^{\alpha m\beta n}}\partial_{\mu}\partial_{\nu}\partial_{\alpha}\partial_{\beta}h_{mn}&=R\indices{^{ambn}}\partial_{\mu}\partial_{\nu}\partial_{a}\partial_{b}h_{mn}+R\indices{^{imjn}}\partial_{\mu}\partial_{\nu}\partial_{i}\partial_{j}h_{mn} \nonumber \\
	=&-\frac{f'}{2r}g^{ab}g^{mn}\partial_{\mu}\partial_{\nu}\partial_{a}\partial_{b}h_{mn}+\frac{1-f}{r^{2}}\left(g^{mn}g^{ij}-g^{mj}g^{ni}\right)\partial_{\mu}\partial_{\nu}\partial_{i}\partial_{j}h_{mn} \nonumber \\
	=&0\,,
\end{align}
where we used the fact that the tensor perturbations are transverse and traceless.
This implies that we have to go to dimension-8 operators to find our first explicit example of higher-derivative term.

\subsection{Dimension-8 operators}
\label{subapp: dim 8 ops}
We now show how that next-order EFT contribution in \eqref{eq: S8 extension} does indeed lead to genuine higher-derivative terms in the equations of motion. We denote the background equations of motion derived from this operator as $C^{\mu\nu}$, and the associated perturbation equations of motion as $\delta C^{\mu\nu}$.

Once again, we want to find terms that lead to genuine higher-derivative terms in the perturbation equations of motion. The only term with four derivatives acting on the metric perturbation for tensor modes is:
\begin{equation}
    \delta C^{\mu\nu} = -16 \mu^{3} c_{\text{R4}} r_{g}^{6}R\indices{^{\alpha\rho\sigma\mu}}R\indices{^{\beta\kappa\lambda\nu}}\nabla_{\rho}\nabla_{\sigma}\nabla_{\kappa}\nabla_{\lambda}h_{\alpha\beta} + \ldots,
\end{equation}
where the ellipses stand for terms that are trivially lower-derivative terms for reasons similar to those in the last two subsections.

Since the tensor perturbations only have non-vanishing components in the angular directions, we pick $(\mu,\nu)=(i,j)$. 
Once again, we are interested in terms with the highest numbers of derivatives acting on the tensor mode, so we trade covariant for partial derivatives.
Then, making use of the lower-order equations of motion $\Box\approx 0$ as well as equation \eqref{eq: Riemann components} for the Riemann tensor components, we find:
\ba
    && R^{m\rho\sigma i}R^{n\kappa\lambda j}\partial_{\rho}\partial_{\sigma}\partial_{\kappa}\partial_{\lambda}h_{mn} \nonumber \\
    &&=\left[\left(\frac{f'}{2r}\right)^{2}\partial_{a}\partial^{a}\partial_{b}\partial^{b}h^{ij}-\frac{(1-f)f'}{r^{3}}\partial_{a}\partial^{a}\partial_{m}\partial^{m}h^{ij}+\left(\frac{1-f}{r^{2}}\right)^{2}\partial_{m}\partial^{m}\partial_{n}\partial^{n}h^{ij}\right] \nonumber \\
    &&=\frac{r_{g}^{2D-6}}{r^{2D-2}}\left[\frac{(D-3)^{2}}{4}\partial_{m}\partial^{m}\partial_{n}\partial^{n}h^{ij}+(D-3)\partial_{n}\partial^{n}\partial_{m}\partial^{m}h^{ij}+\partial_{m}\partial^{m}\partial_{n}\partial^{n}h^{ij}\right] +\mathcal{O}\left(\mu\right) \nonumber \\
&&=\frac{r_{g}^{2D-6}}{r^{2D+2}}\frac{(D-1)^{2}}{4}\LB^{2}h^{ij}+\mathcal{O}\left(\mu\right).
\ea
Including the lower-order terms presents in the action, the full perturbation equation of motion is therefore of the form
\begin{align}
   0=&\frac{1}{2}\delta G_{ij}+\delta B_{ij}+\delta C_{ij}  \\
    =&-\frac{1}{4}\Box h_{ij}+\mu \cGB \left(\ldots\right)+\mu^{2} \cGB^{2} \left(\ldots\right)-4(D-1)^{2} c_{\text{R4}} \mu^{3} \frac{r_{g}^{2D}}{r^{2D+2}}k_{T}^{4}\left(1+\mathcal{O}\left(k_{T}^{-2}\right)\right)h_{ij}\,,\nn
\end{align}
where the $\mathcal{O}\left(\mu\right)$ terms are known from before, and we have not computed the lower-derivative corrections at $\mathcal{O}\left(\mu^{2}\right)$ or $\mathcal{O}\left(\mu^{3}, k_{T}^{2}\right)$.

Now, recall that the master equation is of the form
\begin{equation}
    \Box_{2}\Phi-\frac{V}{f}\Phi=0.
\end{equation}
The metric perturbations are related to the master variable simply by a rescaling.
We can therefore just read off the $\mathcal{O}\left(\mu^{3}\right)$ correction to the effective potential:
\begin{equation}
    \frac{V_T^{\rm D8}}{f} = -16(D-1)^{2} c_{\text{R4}} \mu^3 \frac{1}{r^2} \rgr^{2D} k_T^4 \left(1+ \mathcal{O}\left(k_{T}^{-2}\right)\right).
\end{equation}

\section{Galileon perturbations}
\label{sec: Galileon appendix}

The effective $W_{\ell}$ \eqref{eq: master eq tortoise} associated with the action \eqref{eq: galileon} for the background \eqref{eq: galileon back} is given by:
\ba\label{eq: Galileon Wl}
    W_{\ell}&=&e^{2\rho}\omega^{2}-b^{2}\omega^{2} \\
    &-&\frac{3c_{1}\alpha^{2}}{\Lambda^6}(D-3)^{2}\big[4e^{2\rho}\omega^{2}(D-2)^{2}+((D-1)(3D-5)-4b^{2}\omega^{4})(D-1)(D-3)\big]e^{-2(D-1)\rho} \nonumber \\
    &-&\frac{d_{1} \alpha^2}{4\Lambda^8}(D-3)^{2}\big[-16e^{4\rho}\omega^{4}(D-2)^{2}+8e^{2\rho}\omega^{2}(4b^{2}\omega^{2}+15D^{2}-40D+13)(D-2)(D-1) \nonumber \\
    &&\quad\quad -16b^{4}\omega^{4}(D-1)^{2}-8b^{2}\omega^{2}(D-1)(15D^{3}-67D^{2}+71D+5) \nonumber \\
    &&\quad\quad + (D-1)(127D^5-567 D^4+610 D^3+46 D^2-33D+9)
    \big]e^{-2D\rho} \nonumber \\
    &-&\frac{6d_{2}\alpha^2}{\Lambda^8}(D-3)^{3}(D-2)(D-1)\big[e^{2\rho}\omega^{2}(2D-1)+2D(-b^{2}\omega^{2}+D^{2})\big]
    e^{-2D\rho} \nonumber \\
    &-&\frac{d_{4}\alpha^2}{2\Lambda^8}(D-3)^{2}\big[-16e^{4\rho}\omega^{4}(D-2)^{2}+8e^{2\rho}\omega^{2}(4b^{2}\omega^{2}+7D^{2}-12D+1)(D-1)(D-2) \nonumber \\
    &&\quad\quad -16b^{4}\omega^{4}(D-1)^{2}-8b^{2}\omega^{2}(D-1)(7D^{3}-27D^{2}+23D+5) \nonumber \\
    &&\quad\quad+ (D-1)(63D^5-247 D^4+226 D^3+46 D^2-33D+9)\big]e^{-2D\rho}  \, .\nn
\ea


\bibliographystyle{JHEP}
\input{GB_CausalityD_30.bbl}

\end{document}

%% file: GB_CausalityD_30.bbl
\providecommand{\href}[2]{#2}\begingroup\raggedright\endgroup